\documentclass{svjour3}                     
\smartqed  
\usepackage{graphicx}
\usepackage[top=30truemm,bottom=30truemm,left=25truemm,right=25truemm]{geometry}
\usepackage{amsmath}
\usepackage{amsfonts}
\usepackage{bm}
\usepackage{color}
\graphicspath{{./fig/}}

\definecolor{blue}{rgb}{0, 0, 0}
\definecolor{red}{rgb}{0, 0, 0}
\definecolor{green}{rgb}{0.4660, 0.6740, 0.1880}

\newcommand{\fg}[1]{{\color{blue}#1}}

\journalname{Theor. Comput. Fluid. Dyn.}
\begin{document}

\title{Convolutional neural networks for fluid flow analysis: toward effective metamodeling and low-dimensionalization}


\author{Masaki Morimoto \and Kai Fukami \\
Kai Zhang \and Aditya G. Nair \and Koji Fukagata 
}


\institute{Masaki Morimoto, Koji Fukagata \at
              Department of Mechanical Engineering, Keio University, Yokohama, 223-8522, Japan \\
              \email{masaki.morimoto@kflab.jp}
              \and
              Kai Fukami \at
              Department of Mechanical and Aerospace Engineering, University of California, Los Angeles, CA 90095, USA\\
              Department of Mechanical Engineering, Keio University, Yokohama, 223-8522, Japan
              \and
              Kai Zhang \at
              Department of Mechanical and Aerospace Engineering, Rutgers University, Piscataway, NJ 08854, USA
              \and
              Aditya G. Nair \at
              Department of Mechanical Engineering, University of Nevada, Reno, NV 89557, USA
}

\date{Received: date / Accepted: date}

\maketitle

\begin{abstract}
{
We focus on a convolutional neural network (CNN), which has recently been utilized for fluid flow analyses, from the perspective on the influence of various operations inside \fg{it} by considering some canonical regression problems with fluid flow data.
We consider two types of CNN-based fluid flow analyses; 1. CNN metamodeling and 2. CNN autoencoder.
For the first type of CNN with additional scalar inputs, which is one of the common forms of CNN for fluid flow analysis, we investigate the influence of input placements in the CNN training pipeline.
As an example, estimation of drag and lift coefficients of \fg{an inclined} flat plate and two side-by-side cylinders \fg{in laminar flows is} considered.
For the example of flat plate wake, we use the chord Reynolds number $Re_c$ and the angle of attack $\alpha$ as the additional scalar inputs \fg{to provide} the information \fg{on} the complexity of wake\fg{.}
For the wake interaction problem comprising flows over two side-by-side cylinders, the gap ratio and the diameter ratio are utilized as the additional inputs.
We find that care should be taken for the placement of additional scalar inputs depending on the problem setting and the complexity of flows \fg{that} users handle.
We then \fg{discuss} the influence of various parameters and operations on \fg{the} CNN performance, with the utilization of autoencoder (AE).
A two-dimensional decaying homogeneous isotropic turbulence is considered for the demonstration of AE.
The results obtained through the AE highly rely on the decaying nature.
\fg{I}nvestigation \fg{on} the influence of padding operation at a convolutional layer is also performed.
The zero padding shows reasonable ability compared to other methods which account for \fg{the} boundary conditions \fg{assumed in the}
numerical data.
Moreover, the effect of the dimensional reduction/extension methods inside CNN is also examined.
The CNN model is robust \fg{against} the \fg{difference in} dimension reduction operations, while it is sensitive to the dimensional extension methods.
The findings \fg{of this} paper \fg{will} help us \fg{better design a CNN architecture for practical fluid flow analysis.}
}
\end{abstract}
\section{Introduction}

\begin{figure*}
    \centering
		\includegraphics[width=0.98\textwidth]{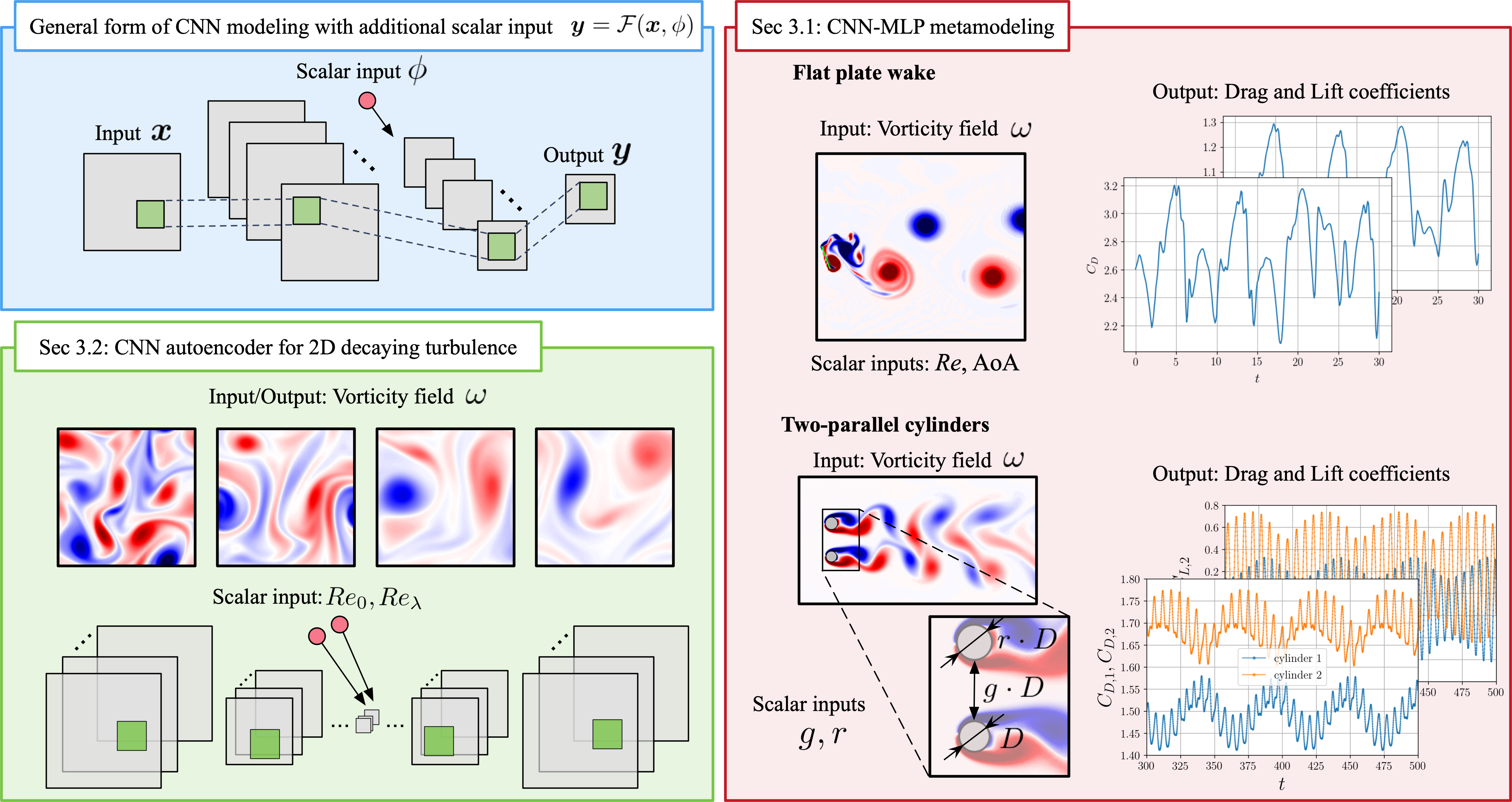}
		\caption{
		Overview of the present study.
		We consider a general form of CNN-based modeling with additional scalar input ${\bm y}={\cal F}({\bm x},{\phi})$.}
		\label{fig1}
\end{figure*}

Convolutional neural networks (CNNs) have recently been recognized as a powerful tool to analyze complex fluid flow phenomena~\cite{FFT2020}.
Although they are still in the phase of fundamental studies, the great potential of CNN-based analyses can be found in turbulence modeling~\cite{DIX2019}, data reconstruction~\cite{BEF2019,nakamura2021MLLSE}, and low-dimensionalization~\cite{BHT2020}.
However, current uses of CNNs require not only the tuning of a bunch of parameters but also the incorporation of {\it a priori} knowledge into its modeling.
Due to these issues, the recent uses of CNNs for fluid flows are based on trial-and-error iterations by users because of a lack of guidelines. 
In this paper, we address the aforementioned issues by focusing on a general form of CNN-based regression tasks for fluid flow analysis ${\bm y}={\cal F}({\bm x},{\phi})$, where ${\bm x}$ and ${\bm y}$ respectively indicate input and output data of machine learning model ${\cal F}$ with additional scalar inputs $\phi$.

One of {\color{red}the most active areas for applying} CNN for fluid flow analyses is turbulence modeling~\cite{DIX2019,DK2021}.
Lapeyre et al.~\cite{lapeyre2019training} utilized a U-net-based CNN for estimating sub-grid scale reaction rates in their large-eddy simulation (LES) framework.
At that moment, previous studies regarding NNs and LES sub-grid modeling had been conducted using a multi-layer perceptron (MLP)~\cite{RHW1986}.
Hence, the use of CNN for the LES closure was one of the novelties of their study.
The comparison between the MLP and the CNN for LES closure modeling is well discussed by Pawar et al.~\cite{pawar2020priori}.
The CNN can also be applied to turbulence modeling for Reynolds-averaged Navier--Stokes (RANS) simulations.
{\color{red}Thuerey et al.~\cite{thuerey2020deep} considered CNN-based estimation for the airfoil flow field obtained from RANS and showed that NNs are even applicable to RANS data, dispelling the skepticism on the applications of NNs to RANS.}
In addition to the aforementioned efforts on LES and RANS, Font et al.~\cite{font2020deep} has recently proposed new concept for the decomposition of Navier--Stokes equation called {\it spanwise-averaged N--S equation}.
Analogous to LES and RANS equations, it also has a closure term.
They used a CNN-based regressor to close that term with a velocity field while considering the example of wakes behind a cylinder and an ellipse.

CNN-based fluid flow data reconstruction can also be regarded as a promising field.
For instance, the CNN-based super-resolution analysis proposed by Fukami et al.~\cite{FFT2019a} is a proof of concept of fluid flow data recovery with machine learning.
The aim of the super-resolution analysis is to reconstruct high-resolution data from its low-resolution counterpart.
Since this low-resolution part can also be replaced \fg{by} various forms \fg{of} low-resolution fluid flow information, e.g., local sensor measurements and limited availability of data, the super-resolution idea has been extended to not only numerical~\cite{LTHL2020,kim2020deep} but also experimental studies~\cite{DHLK2019,MFF2020,CZXG2019}.
From the view of estimation, the combination of CNN and MLP can widely be seen for their purposes, since their output shape \fg{is} often \fg{in} a form of scalar.
For example, Salehipour and Peltier~\cite{SP2019} attempted to predict the small-scale structures in the ocean turbulence called {\it atoms} using a CNN.
Otherwise, Morimoto et al.~\cite{MFZF2020} visualized the internal procedures of the CNN-MLP model with fluid flow analyses toward practical applications from the perspective on interpretability.

Furthermore, the CNN-based modeling is also capable of low-dimensionaliz\fg{ing} fluid flows, as a form of autoencoder (AE)~\cite{MFF2019}.
The AE has the same output data as the input while having \fg{a dimension}
compression procedure inside its structure.
Because the AE is trained to output the same data as the input, the latent variables, which can be extracted from the bottleneck layer inside the AE, can be regarded as a low-dimensionalized representation of high-dimensional data, if the reconstruction via the AE is well performed. 
To the best of our knowledge, Milano and Koumoutsakos~\cite{Milano2002} first brought the idea of AE into the fluid dynamics field.
Although their model was based on the MLP, the CNN has also recently been applied to build AE thanks to the CNN's great advantage against the MLP in terms of the number of weights inside the model while being able to keep its accuracy~\cite{LKB2018,FNF2020}.
In addition, the extracted low-dimensional representations via AE can also be utilized for the construction of reduced-order modeling which has a similar form \fg{to that} of \fg{the} proper orthogonal decomposition-based Galerkin integration~\cite{HFMF2019,MLB2021,fukami2020sparse,CJKMPW2019,FNKF2019}.

As discussed above, we are now able to appreciate the strong potential and applicability of CNN for fluid flow analyses.
However, the current success of CNN and fluid flow analysis is based on trial-and-error iterations by fluid mechanicians since we have no guideline for parameter decisions inside the CNN.
For instance, the work on super-resolution analysis introduced above~\cite{FFT2019a} reported the importance of the utilization of multi-size filters inside CNNs so as to account for a wide range of scales included in turbulence.
A similar idea can also be found in the construction of NN-based reduced order modeling~\cite{nakamura2020extension,LPBK2020} and surrogate models for high-fidelity simulations~\cite{LY2019}.
Otherwise, several reports utilize additional scalar inputs which highly relates to fluid flow phenomena, e.g., angle of attack, Reynolds number, and bluff body shapes, to improve the estimation or low-dimensionalization abilities of CNNs~\cite{LKT2016,HFMF2020a,HFMF2020b,zhang2018application,bhatnagar2019prediction,kim2020deep,xu2020multi,ZXCLCPV2020,PSARK2020}.
Hence, we now arrive at the question: {\it how can we determine these strategies to achieve a better performance of CNN and fluid flows?} --- In this paper, we tackle this vague portion with regard to the use of CNN and fluid flow analyses.

As mentioned earlier, we consider CNN-based regression tasks ${\bm y}={\cal F}({\bm x},{\phi})$ of fluid flow analyses.
In this paper, we consider two types of CNN-aided modeling, as illustrated in figure~\ref{fig1}: 1. CNN-MLP based metamodeling of unsteady laminar wakes and 2. CNN-AE for turbulence, so as to investigate various considerable parameters inside CNN.
The organization of the present paper is as follows: 
We first introduce the basic principle of CNN and the covered models based on CNN in section~\ref{sec:cnnfluids}.
The information for fluid flow data used in the present paper is offered in section~\ref{sec:fluidflowdata}.
Results and discussions are provided in section~\ref{sec:results}.
We finally give concluding remarks in section~\ref{sec:conclusion}.

\section{Methods}
\label{sec2}

\subsection{Convolutional neural network (CNN) for fluid flow analyses}
\label{sec:cnnfluids}

\subsubsection{Basic principle of CNN}
\label{sec:cnn}

\begin{figure*}
    \centering
		\includegraphics[width=0.75\textwidth]{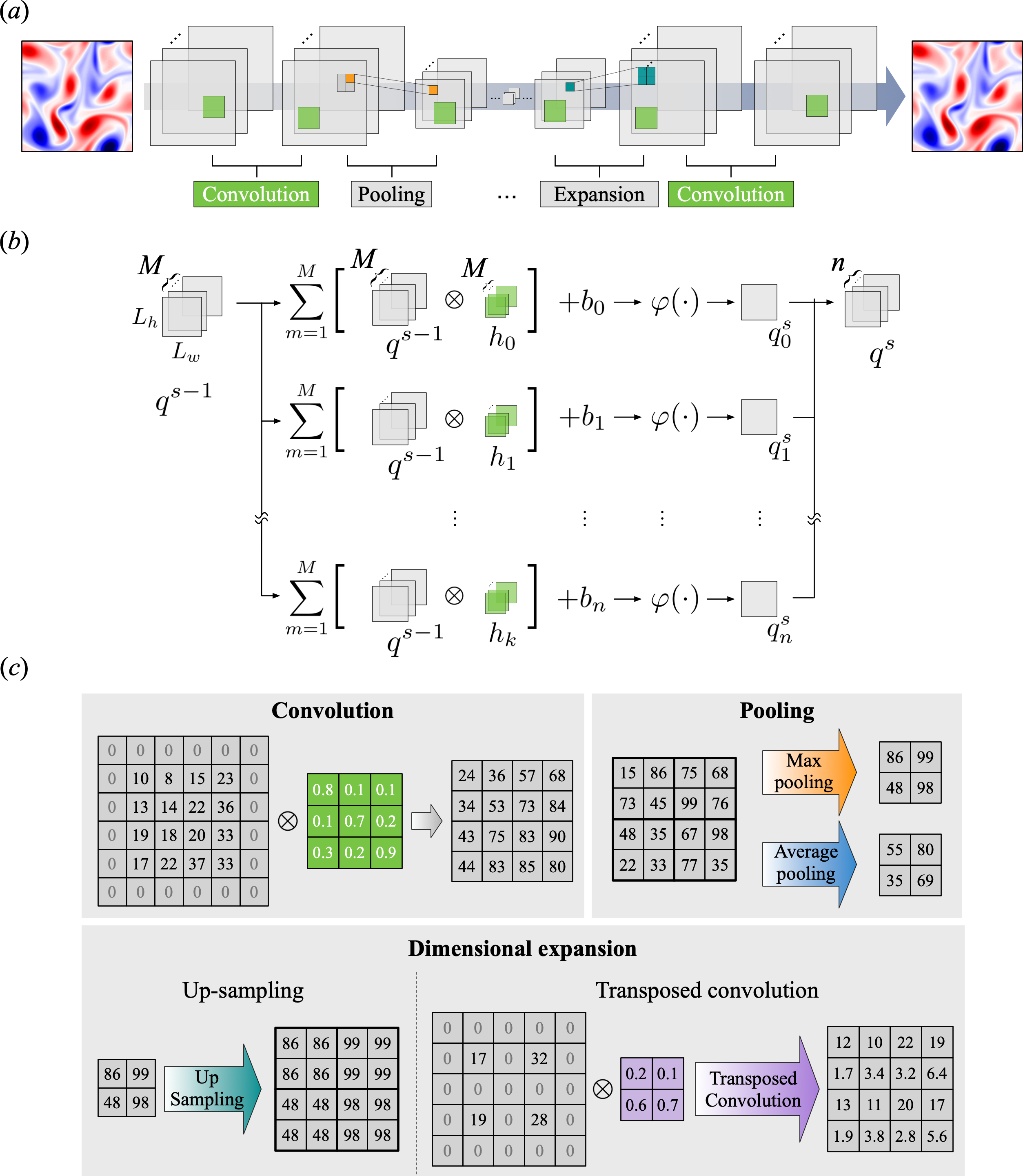}
		\caption{
		Basic operations of CNN.
		$(a)$ A structure of typical CNN-AE with convolutional layers, pooling layers, and dimensional expanding layers.
		$(b)$ Convolutional operation at each convolutional layer to obtain $q^s$ from $q^{s-1}$.
		$(c)$ Convolution, pooling, and dimensional expansion.
		Gray zero values indicate additionally embedded values for each operation.}
		\label{fig:CNN_operation}
\end{figure*}

Let us first introduce operations inside a convolutional neural network (CNN)~\cite{LBBH1998}.
The CNN was originally developed in image recognition tasks.
In particular, various efforts on CNN-based image classification can now be widely seen such as ResNet~\cite{Simonyan2015}, GoogLeNet~\cite{szegedy2015going}, and SENet~\cite{hu2018squeeze}.
Because CNN is good at handling high-dimensional data through the concept of weight sharing, it is also becoming one of the promising tools to analyze fluid flows~\cite{KL2020,HLC2019,GKO2020,matsuo2021supervised,2021wallmodel}.

A CNN typically consists of several types of layers, i.e., convolutional layer, pooling layer, and up-sampling layer, as illustrated in figure~\ref{fig:CNN_operation}$(a)$ with an example of CNN autoencoder.
A main operation of CNN is performed at the convolutional layer which extracts spatial features of input data through filter operations, as presented in figure~\ref{fig:CNN_operation}$(b)$.
A fundamental operation of convolutional layer shown in figure~\ref{fig:CNN_operation}$(b)$ is taking a summation of an Hadamard product of an arbitrary portion of input data and a filter $h$.
Output data of a convolutional layer $q^{(s)}$ can be expressed as,
\begin{equation}
    q^{(s)}_{ijn}=\varphi\left(\sum_{m=1}^M\sum_{p=0}^{H-1}\sum_{q=0}^{H-1}h^{(s)}_{pqmn}q^{(s-1)}_{i+p-G,j+q-G,m}+b_n^{(s)}\right),
    \label{eq:CNN}
\end{equation}
where $G=\lfloor H/2\rfloor$ (where $\lfloor\cdot\rfloor$ represents the operation of rounding down the value to the nearest decimal), $H$ is width and height of the filter, $M$ is the number of input channel, $n$ is the number of output channel, $b$ is a bias, and $\varphi$ is an activation function, respectively.
Note that we showed the two-dimensional operation above since two-dimensional flows are only handled through the present study, although its extension to three-dimensional flows is rather straightforward~\cite{FFT2021b,nakamura2020extension} albeit computationally more expensive.
The nonlinear activation function $\varphi$ enables a machine learning model to account for nonlinearlities into its estimation.
There are various choices of nonlinear activation functions~\cite{FHNMF2020}.
We utilize the ReLU function~\cite{NH2010} which can avoid vanishing the gradient of weights in deep CNNs.
Weights ${\bm w}$ (values on filters) are optimized through a backpropagation~\cite{Kingma2014} to minimize the loss function between estimated data and reference data ${\bm q}_{\rm Ref}$,
\begin{equation}
    {\bm w}={\rm argmin}_{\bm w}||{\bm q}^{(s_{\rm max})}-{\bm q}_{\rm Ref}||_2,
    \label{eq2}
\end{equation}
where ${\bm q}^{(s_{\rm max})}$ is an output of CNN at the last layer $s_{\rm max}$.

We also incorporate other layers to process data inside \fg{the} CNN.
One of them is a pooling layer which downscales the data.
This feature is useful \fg{in reducing data dimension} for regression or classification problems --- for example, to estimate a scalar value from high-dimensional data~\cite{FFT2020}, and to extract key features using an autoencoder~\cite{HS2006,MFF2019,FNF2020}.
There are mainly two methods for downsampling in the pooling layer, i.e., max and average pooling, as illustrated in figure~\ref{fig:CNN_operation}$(c)$.
Both methods are common in extracting a representative value from an arbitrary area; the difference only lies in whether to extract, max, or average values.
This difference can mathematically be written as,
\begin{equation}
q_{ij}^{\rm LR}=\left(\frac{1}{\gamma^2}{\sum_{k,l\in P_{i,j}}(q_{kl}^{\rm HR})^P}\right)^{1/P},
\end{equation}
where $P=1$ or $\infty$ provides average or max pooling, for the arbitrary region ($\gamma\times\gamma$) of the input data $q^{\rm HR}$.

Contrary to the pooling operation, we can also consider expanding the dimension of the data\fg{, which} is required \fg{,e.g.,} in constructing a CNN autoencoder (for decoder part) and in estimating the two-dimensional sectional flow field from scalar values of sensor measurements using CNN~\cite{FFT2020,MFF2019,Milano2002,FHNMF2020,PLACW2020}.
There are also mainly two techniques to expand the dimension, i.e., up-sampling and transposed convolution, as shown in figure~\ref{fig:CNN_operation}$(c)$.
The up-sampling is \fg{a} simple operation \fg{which} copies the value onto an arbitrary region.
On the other hand, the transposed convolution~\cite{deconv2015,DV2015} is fundamentally an inverse operation of regular convolution.
As illustrated in figure~\ref{fig:CNN_operation}$(c)$, the input data \fg{are}
first expanded by embedding zero value among grid points.
The regular convolutional operation (equation~\ref{eq:CNN}) is then applied to expand the dimension of data.

As introduced above, we have various candidates to construct a CNN depending on users' tasks. 
{\color{red}However, to the best of our knowledge, the influence of various parameters inside CNNs on their ability has not yet been investigated in detail, despite that these operations are deemed to keys for CNN-based fluid flow analyses.}
In this study, we address this point by focusing on the choice of downsampling operations and dimensional expanding techniques.

\subsubsection{Covered CNN models}
\label{sec:covered_cnn}
\begin{figure*}
    \centering
		\includegraphics[width=0.75\textwidth]{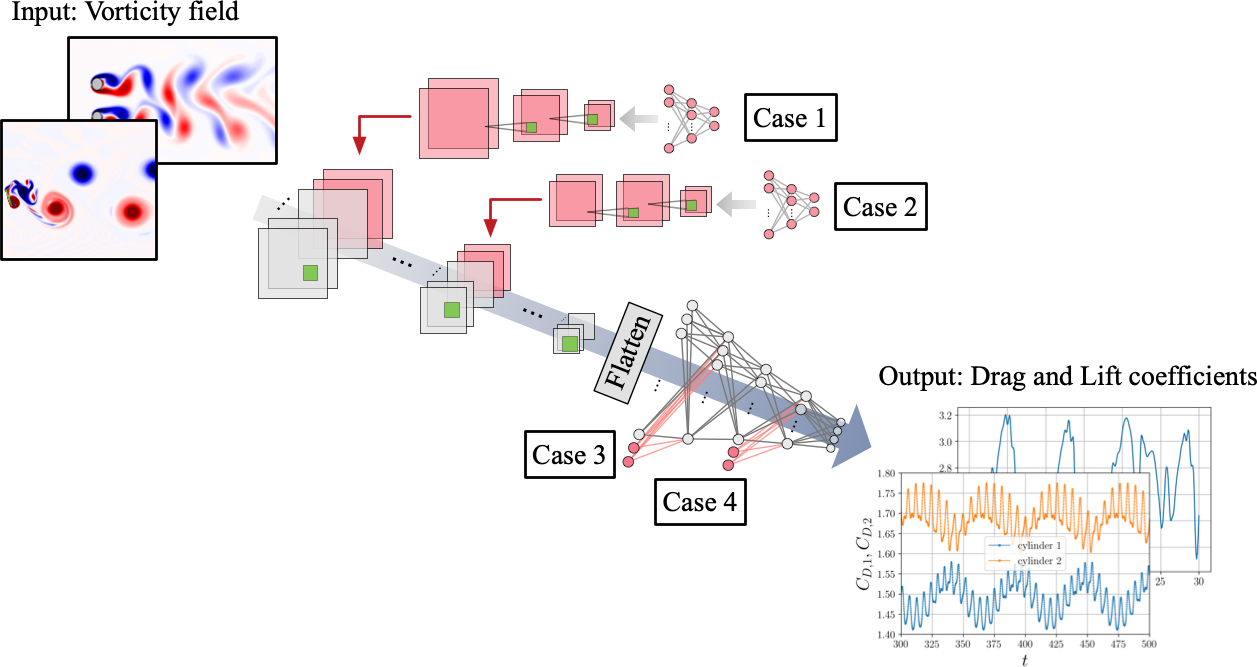}
		\caption{
		CNN-MLP model covered in this study.
		Pink nodes indicate additional scalar inputs of the flow; 1. $\{r,g\}$ for two side-by-side cylinders flow and 2. $\{Re_c,\alpha\}$ for flat plate wake.}
		\label{fig:CNN-MLP}
\end{figure*}
\begin{figure*}
    \centering
		\includegraphics[width=0.65\textwidth]{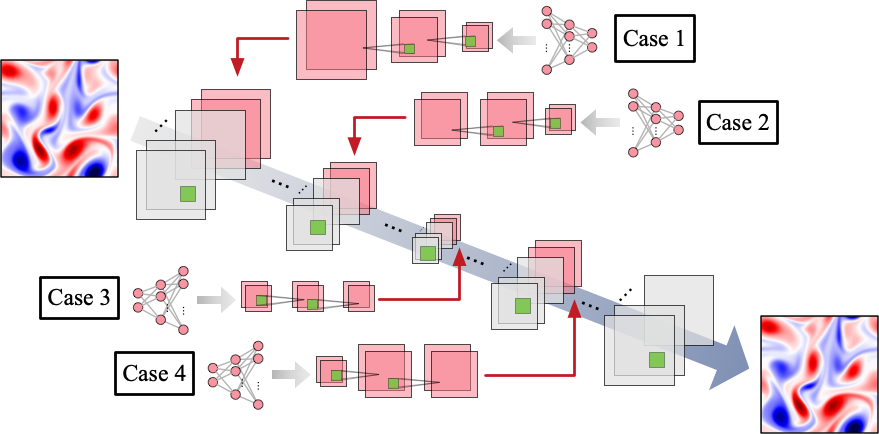}
		\caption{
		CNN-AE model utilized with two-dimensional decaying turbulence.
		Pink nodes indicate additional scalar inputs of the flow, $\{Re_0,Re_{\lambda}\}$.}
		\label{fig:CNN-AE}
\end{figure*}

As mentioned above, we consider two types of convolutional neural network (CNN)-based architectures: 1. combination of CNN and multi-layer perceptron (CNN-MLP) and 2. CNN-based autoencoder (CNN-AE), which have widely been utilized in both regression and classification tasks~\cite{LBH2015}.

Most of the image classification models consist of the CNN-MLP model since it generally outputs scalar values as a probability of each class from sectional or volumetric data~\cite{VGG16}.
It is also utilized in fluid flow analyses when it \fg{is required} to output scalar variables, e.g., aerodynamics coefficients, from two-dimensional data such as flow fields~\cite{zhang2018application}.
\fg{Considering these, the present CNN-MLP model is constructed as follows;}
\begin{enumerate}
    \item Convolutional layers and pooling layers are respectively utilized to extract key features and to downsample images, as shown in figure~\ref{fig:CNN-MLP}.
    \item An MLP is then adopted to output scalar values after reshaping (i.e., `Flatten' in figure~\ref{fig:CNN-MLP}) the data to a one-column matrix.
\end{enumerate}
In this study, we aim to estimate drag and lift coefficients of flows over two side-by-side cylinders and a flat plate.

When utilizing CNNs for coefficient estimations of fluid flows, additional scalar values which strongly relates to fluid flow phenomena, e.g., the Reynolds number, the Mach number, and the angle of attack, are often inserted to help its estimations~\cite{zhang2018application,bhatnagar2019prediction}.
{\color{red}Although this is a popular technique, the placement of these scalar inputs has usually been decided by users without thorough investigation to date.}
To investigate this point, we consider additional scalar inputs for each example,
\begin{enumerate}
    \item the chord Reynolds number $Re_c$ and the angle of attack $\alpha$ for a flat plate wake, and
    \item the diameter ratio of two cylinders $r$ and distance between two cylinders $g$ for two parallel cylinders' wake,
\end{enumerate}
from four different cases of placements, as illustrated in figure~\ref{fig:CNN-MLP}.
For cases 1 and 2, the dimension of scalar inputs is expanded using MLP and CNN to concatenate with the output of the first and third convolutional layers.
In contrast, the scalar values are directly concatenated to the first and sixth layers of MLP for cases 3 and 4.

Another well-known method of CNN-based fluid flow analyses is the utilization as \fg{an autoencoder (AE).}
An AE~\cite{HS2006} is used to map high-dimensional data into \fg{a} low-dimensional feature space referred to as \fg{the} latent vector.
In particular, CNN-AE, an AE with convolutional layers, is known as a good candidate to extract key features of two-dimensional images~\cite{MFF2019,omata2019,HFMF2020a,HFMF2020b,nakamura2020extension}.
An encoder part --- the first half of CNN-AE --- contains the pooling layers and the convolutional layers to extract key features of input data while reducing the dimension of data.
A decoder part --- the latter half of CNN-AE --- then operates for expanding the dimension of the data.
If we can obtain the same output as the input data through the bottleneck procedure, it implies that the information of input data can successfully be low-dimensionalized into the latent space.
In the present study, the up-sampling layer and the transposed convolutional layer are considered for the means of dimension expansion, as introduced in section~\ref{sec:cnn}.

For the investigation of CNN-AE, we utilize a decaying homogeneous isotropic turbulence, as presented in figure~\ref{fig:CNN-AE}.
To concatenate the scalar values with the convolutional layers, these scalars are expanded to two-dimensional sectional data using the MLP and convolutional layers similar to the investigation of the CNN-MLP model.
Here, we consider four different input placements, i.e., the input layer (case 1), the third convolutional layer (case 2), the latent space (case 3), and the 16th layer in the decoder part (case 4).

We perform a three-fold cross validation~\cite{BK2019} for all neural networks and utilize the mean $L_2$ error norm to assess the ability of networks for each case.
We use the Adam optimizer~\cite{Kingma2014} for updating the weights,
and training/validation data are randomly sampled for training.
In what follows, the error assessment of CNN-MLP model is performed using test data excluded from the training data process, although the flow configuration is the same as that for training.
For the autoencoder, i.e., CNN-AE, in contrast, we sample test data from the training data range.

\subsection{Fluid flow data sets}
\label{sec:fluidflowdata}

\subsubsection{Flat plate wake}
\begin{figure}
    \centering
    \includegraphics[scale=0.5]{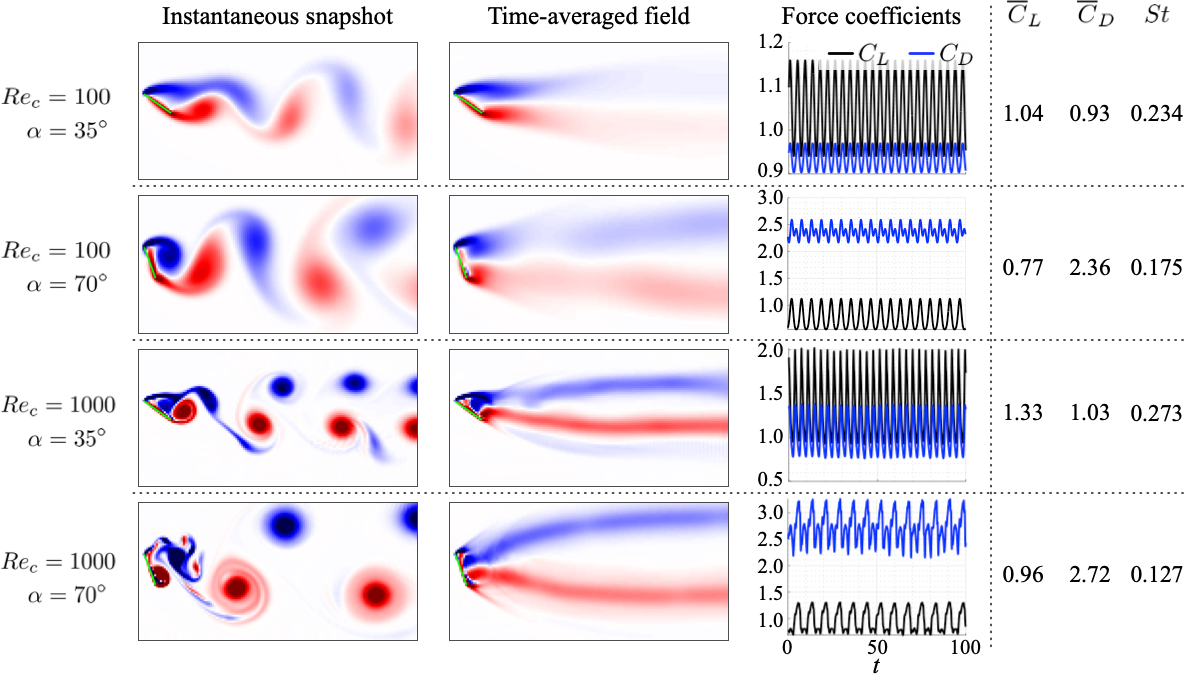}
    \caption{
    Wakes of flow over a flat plate.
    Shown are the instantaneous (first column) and time-averaged (middle column) vorticity fields $\omega\in[-7,7]$, with blue and red representing negative and positive values, respectively.
    The lift and drag coefficients for each case are shown in the last column.}
    \label{fig:FPFlowfields}
\end{figure}

We consider a flat plate wake at the chord-based Reynolds number of $Re_c=\{100,1000\}$ with the angle of attack $\alpha=\{35^\circ,70^\circ\}$.
A two-dimensional direct numerical simulation (DNS) with an immersed boundary projection method~\cite{TC2007} is performed to simulate the flows.
The solver employs a discrete vorticity-stream function formulation with an immersed boundary framework to generate the plate.
The leading edge of the flat plate is placed at $(x/c, y/c) = (0,0)$.
A multi-domain technique~\cite{CT2008} with five grid levels is utilized with the finest inner-most domain as $-0.2 \le x/c \le 1.8, -1 \le y/c \le 1$ and grid spacing of $\Delta x/c = 0.008$.
The solver uses the Crank-Nicolson scheme for the viscous term and an explicit second-order Adam-Bashforth method for the advective term.
At the far-field boundaries, uniform flow is prescribed.
To possess sufficient spatial and temporal resolution for estimation of lift and drag forces, we collect $1200$ snapshots within the spatial domain $-2.5 \le x/c \le 8.73, -4 \le y/c \le 4$ at a sampling frequency of $fU_\infty/c = 10$.

In all four cases analyzed, unsteady vortex shedding behavior is observed in the wake.
The complexity of the vortical structures increases with the angle of attack and also with the Reynolds number, as shown in figure~\ref{fig:FPFlowfields}.
As the nonlinear interactions in the wake increase, the time-averaged flow fields reveal the deviation of the vortical structures from the centerline.
The time-averaged lift $\bar{C}_L$ and drag forces $\bar{C}_D$ along with the dominant shedding frequency $St$ for each case is also summarized in figure~\ref{fig:FPFlowfields}.
The drag forces increase dramatically with the angle of attack while the shedding frequency decreases.

\begin{figure}
    \centering
    \includegraphics[scale=0.5]{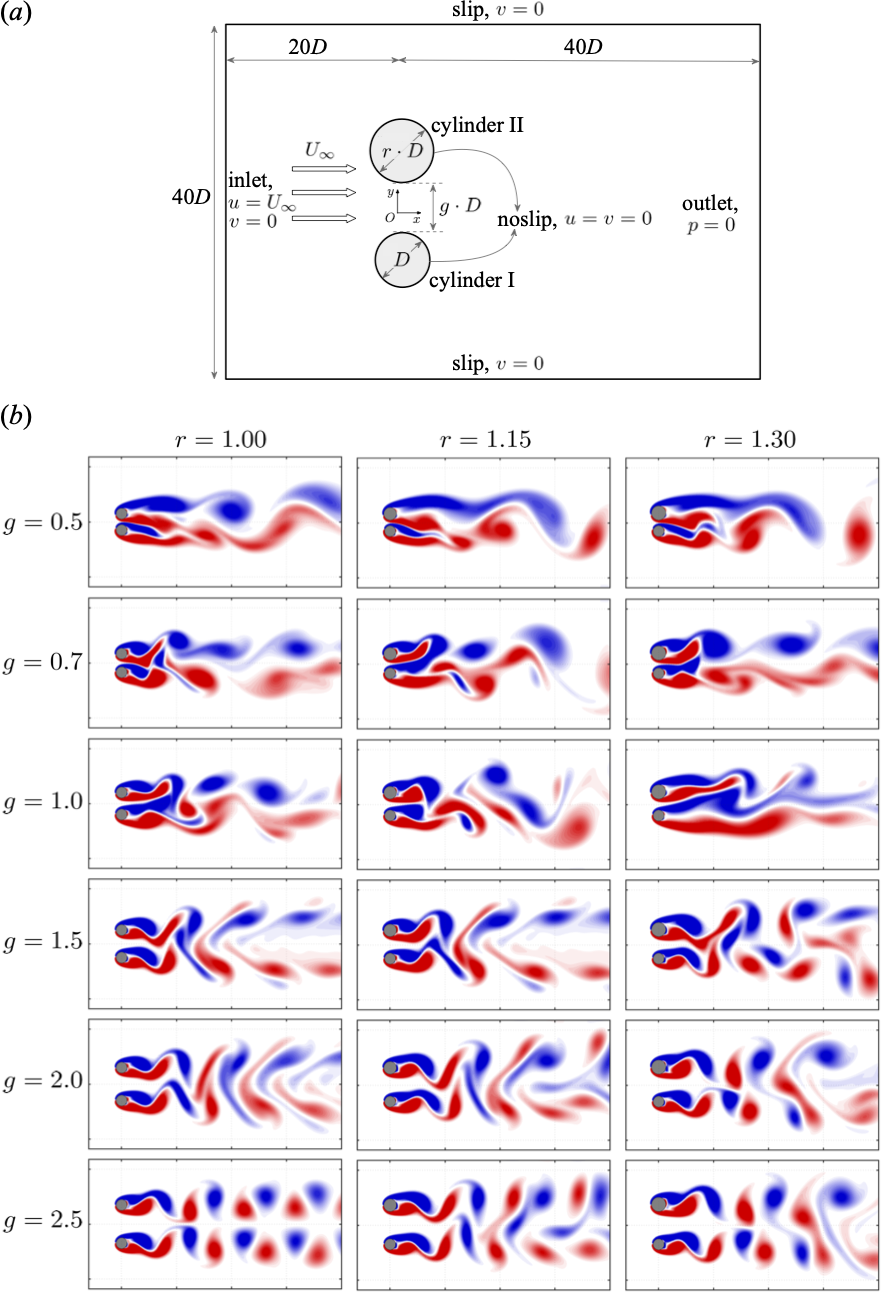}
    \caption{
    $(a)$ Computational domain for flow over two side-by-side cylinders.
    $(b)$ Wakes of flow over two parallel cylinders.
    Shown are the vorticity fields $\omega\in[-1,1]$, with blue and red representing negative and positive values, respectively.}
    \label{fig:2Pflowfields}
\end{figure}
\subsubsection{Wake of two side-by-side cylinders}

The wake interactions between two side-by-side circular cylinders with uneven diameter is considered. 
A schematic view of the problem setup is shown in figure~\ref{fig:2Pflowfields}$(a)$. 
The two circular cylinders with a size ratio of $r$ are separated with a gap of $gD$, where $g$ is the gap ratio. 
The Reynolds number is fixed at $Re_D=U_{\infty}D/\nu=100$.
The two cylinders are placed $20D$ downstream of the inlet where a uniform flow with velocity $U_{\infty}$ is prescribed, and $40D$ upstream of the outlet with zero pressure. 
The side boundaries are specified as slip and are $40D$ apart. 
The flows over the two cylinders are solved by the open-source CFD toolbox OpenFOAM~\cite{weller1998tensorial}, using second-order discretization schemes in both time and space.

The flow physics is governed by two parameters; the size ratio $r$ and the gap ratio $g$. 
The flow fields of $r=\{1.00, 1.15, 1.30\}$
and $g=\{0.5-2.5\}$ are shown in figure \ref{fig:2Pflowfields}. 
For $g=\{0.5, 0.7, 1.0\}$, the wakes are generally chaotic for all three size ratios. 
For higher gap ratios, the wakes restore order, but are characterized by different features. 
In the case of two identical cylinders ($r=1.00$), for $g=\{1.5, 2.0\}$, the vortices shed from the two cylinders are in phase with each other in the near wake, and merge into a larger binary vortex street downstream.
At $g=2.5$, the parallel vortex streets are out-of-phase with each other, \fg{and} they form a pair of symmetric flow patterns with respect to $x$ axis.
For cylinders with a different but close diameter ($r=1.15$), the two vortex streets with different natural shedding frequencies can synchronize to form a single binary vortex street, as is the case for $g=1.5$.
With the increase in the gap ratio, the coupling between the two vortex streets becomes weaker.
As a result, the wakes of $g=\{2.0, 2.5\}$ are characterized by quasi-periodicity due to the interactions between the two vortex streets with different shedding frequencies.
Such quasi-periodic flow also prevails for cases with $(r,g)=(1.30, \{1.5-2.5\})$, for which synchronization does not occur due to the large difference in the natural shedding frequencies.

\subsubsection{Decaying homogeneous isotropic turbulence}

\begin{figure*}
    \centering
	\includegraphics[width=0.95\textwidth]{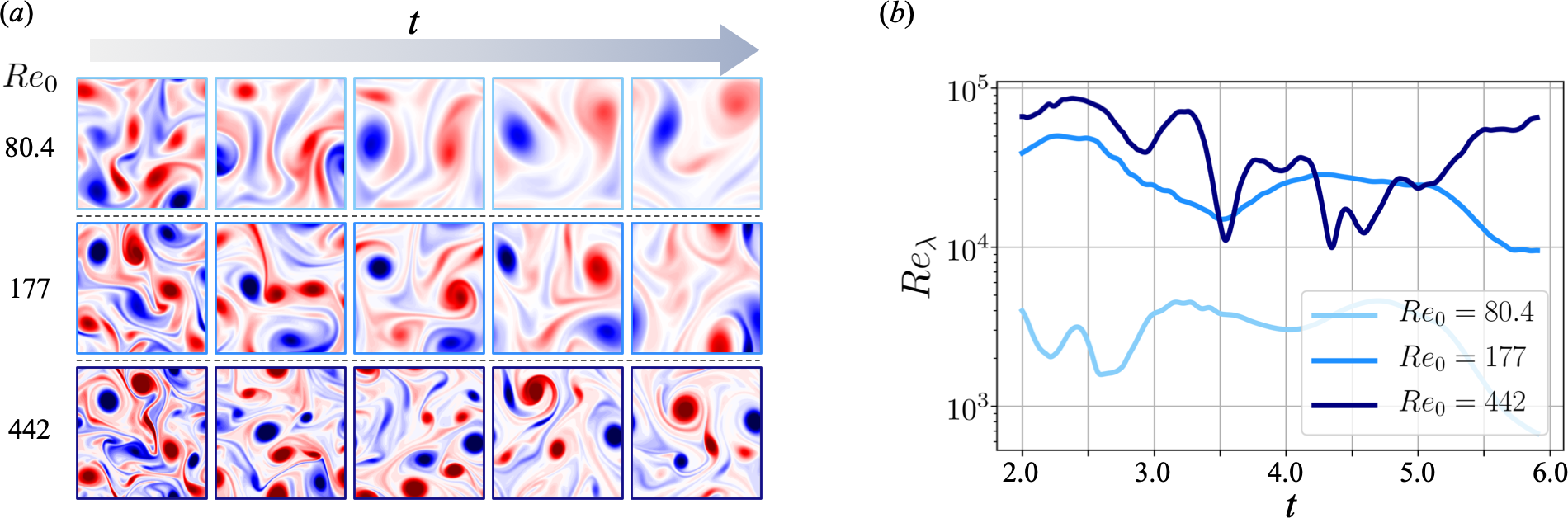}
	\caption{
	Two-dimensional decaying isotropic homogeneous turbulence utilized in this study.
	$(a)$ Vorticity field of flows at each initial Reynolds number $Re_0$ and $(b)$ time evolution of their Taylor Reynolds numbers $Re_\lambda$.}
	\label{fig:turb_varRe}
\end{figure*}

To examine the influence of CNN performance on various parameters inside the CNN-AE, we also consider a two-dimensional decaying homogeneous isotropic turbulence. 
The training data set is prepared by a DNS~\cite{TNB2016}. 
The governing equation is the two-dimensional vorticity transport equation,
\begin{equation}
\partial_t\omega+{\bm u}\cdot\nabla\omega=Re_0^{-1}\nabla^2 \omega,
\label{eq_1}
\end{equation} 
where ${\bm u}=(u,v)$ and $\omega$ are the velocity and vorticity, respectively.  
The size of the computational domain is $L_x=L_y=1$.
In this study, three initial Reynolds numbers $Re_0\equiv u^*l_0^*/\nu=\{80.4,177,442\}$ are considered, where $u^*$ is the characteristic velocity obtained by the square root of the spatially averaged initial kinetic energy, {$l_0^*=[2{\overline{u^2}}(t_0)/{\overline{\omega^2}}(t_0)]^{1/2}$} is the initial integral length, and $\nu$ is the kinematic viscosity, as presented in figure \ref{fig:turb_varRe}.
The numbers of grid points are $N_x=N_y=\{128,256,512\}$ for the covered initial Reynolds numbers $Re_0=\{80.4,177,442\}$, respectively. 
For the input and output attributes to the present AE, the vorticity field $\omega$ is utilized.
Since the size of input data must be consistent over the covered Reynolds numbers to feed into the AE, the vorticity data generated at $Re_0=80.4$ and 442 are interpolated to $256^2$ size when we handle with the AE.
In addition, we also use the instantaneous Taylor Reynolds number $Re_\lambda(t) = u^{\#}(t)\lambda(t)/\nu$, where $u^{\#}(t)$ is the spatial root-mean-square value for velocity of an instantaneous field and $\lambda(t)$ is the Taylor length scale of the instantaneous field, as the second scalar input, as shown in the green portion of figure~\ref{fig1}.
For training the present AE, we use 1000 snapshots for each initial Reynolds number in a dimensionless time of $t=2-6$ with a time interval of $\Delta t=0.004$.

\section{Results and Discussion}
\label{sec:results}

\subsection{Example 1: scalar input-aided convolutional neural networks}
\label{sec:cnn-mlp}
\begin{figure*}
    \centering
	\includegraphics[width=0.89\textwidth]{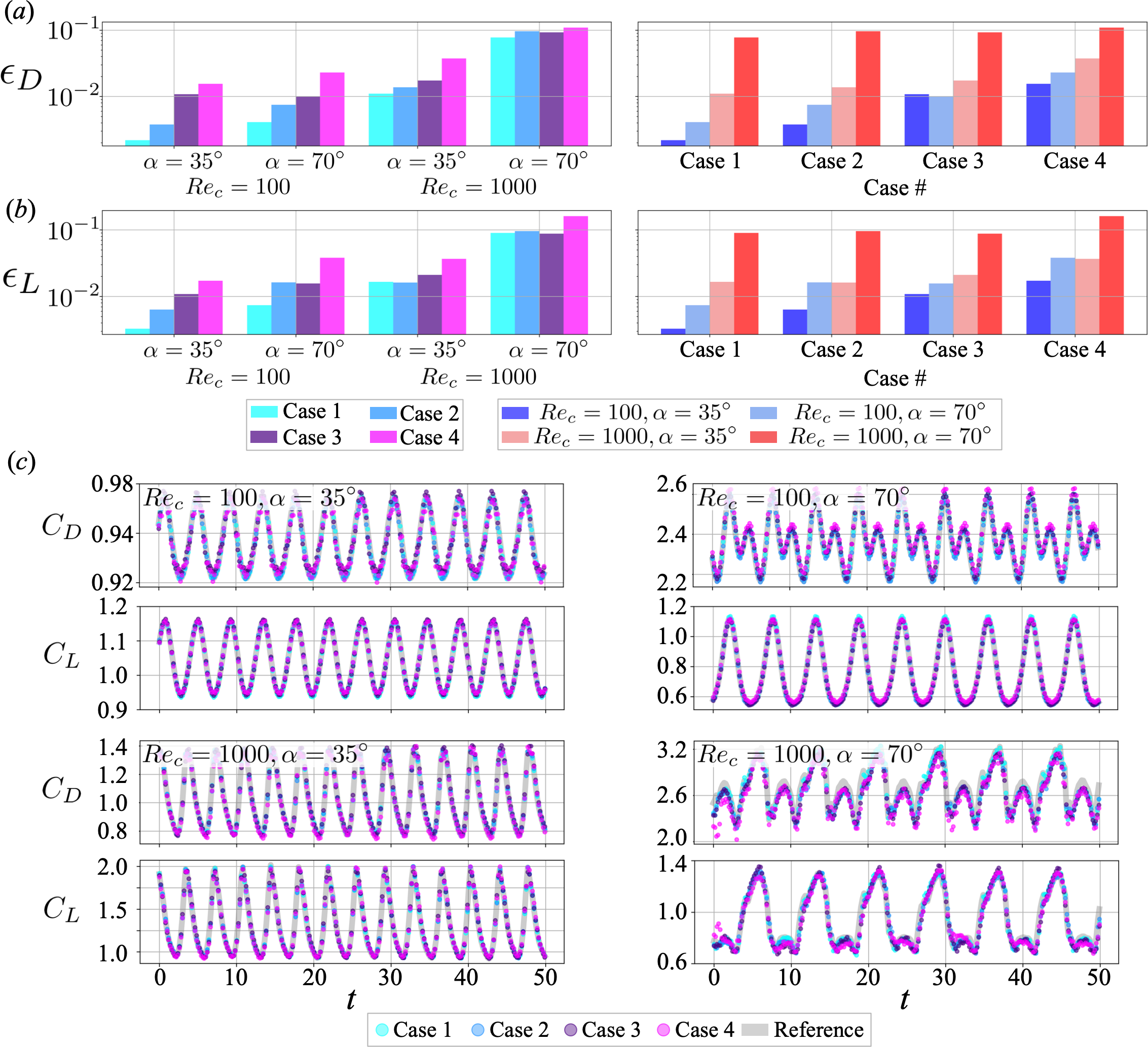}
	\caption{
	Dependence of $C_D$ and $C_L$ estimations on input scalar placements for a flow over a flat plate.
	Comparison of the $L_2$ error norm among flow types (right side) and input placements (left side) with $\{Re_c,\alpha\}$ inputs for $(a)$ $C_D$ and $(b)$ $C_L$.
	$(c)$ Time traces of estimated coefficients.}
	\label{fig:flatplate_result}
\end{figure*}
\begin{figure*}
    \centering
	\includegraphics[width=0.58\textwidth]{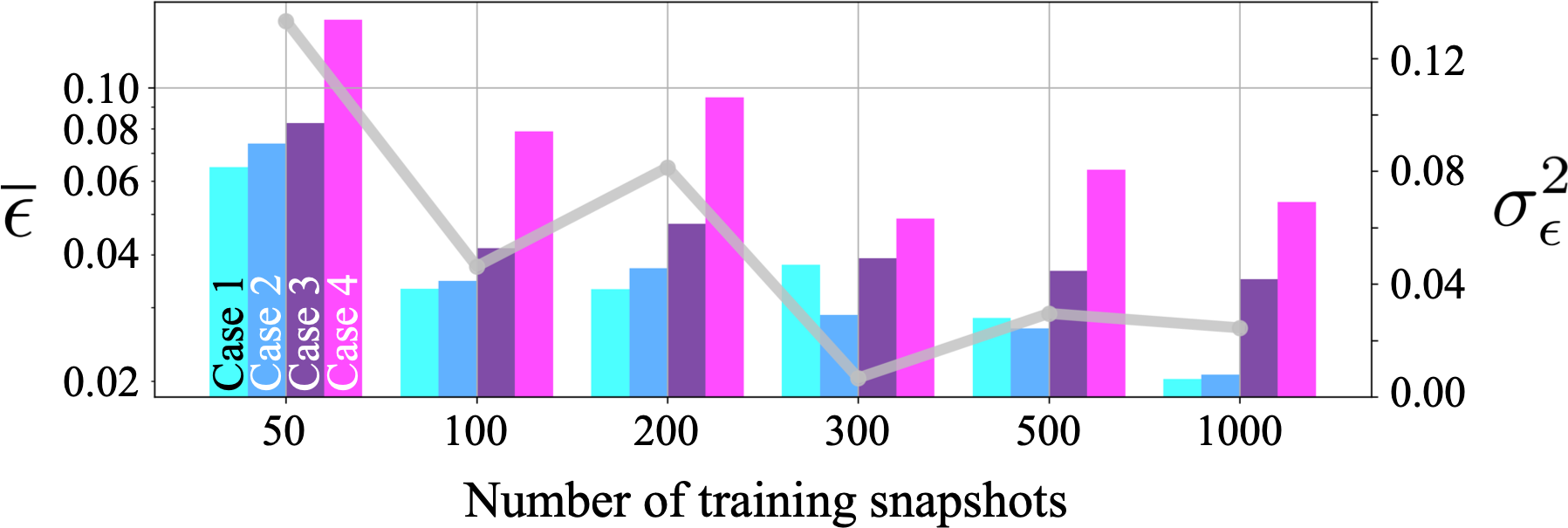}
	\caption{
	Dependence of the averaged $L_2$ error norms $\overline{\epsilon}$ of force coefficient estimations for a flow over a flat plate with $\{Re_c, \alpha\}=\{1000,70^\circ\}$ on the number of training snapshots.
	The standard deviation over the cross validation is also shown as the gray line.}
	\label{fig:flat-plate_nos}
\end{figure*}

We first assess the influence \fg{of the} scalar input placements for the CNN-MLP model.
As stated in section~\ref{sec:covered_cnn}, our consideration is the estimation of drag $C_D$ and lift $C_L$ coefficients of the flow over a flat plate and two side-by-side cylinders.
As already presented in figure~\ref{fig:CNN-MLP}, we investigate four input placements for feeding input scalar values, referred to as cases 1, 2, 3 and 4.
In the present formulation, the dimension of scalar inputs are expanded to concatenate with convolutional layers with cases 1 and 2, while they are directly connected to the MLP layers for cases 3 and 4.
With both flow examples, a single machine learning model handles all types of flows.
For instance, a single model is trained with four types of flows (combination among $Re_c=\{100,1000\}$ and $\alpha=\{35^\circ,70^\circ\}$) for the flat plate wake, while other one is trained with eighteen types of flows (combination among $r=\{1.00,1.15,1.30\}$ and $g=\{0.5,0.7,1.0,1.5,2.0,2.5\}$) for the two side-by-side cylinders example.

The first example for the present analysis is the flat plate wake.
The $L_2$ error norms of the estimated drag and lift coefficients are summarized in figures~\ref{fig:flatplate_result}$(a)$ and $(b)$.
Each $L_2$ error norm are defined as follows; $\epsilon_D=||C_{D,{\rm Pred}}-C_{D,{\rm True}}||_2/||C_{D,{\rm True}}||_2$ for drag coefficient and $\epsilon_L=||C_{L,{\rm Pred}}-C_{L,_{\rm True}}||_2/||C_{L,{\rm True}}||_2$ for lift coefficient, respectively.
The left portion shows the relationship among the covered flow types and the error, while the right counterpart compares the cases of input placements.
The basic trend here is that \fg{the estimation} at higher $Re_c$ and larger $\alpha$ \fg{is less accurate}, although the highest $L_2$ error norm 
\fg{is still as low as} 0.1.
The $L_2$ error norms of $C_D$ and $C_L$ are almost at the same level in this case.
It is striking that the cases with scalar inputs at the upperstream layers tend to show a better estimation than those with the scalar inputs at the downstream layers.
This is likely because feeding the scalar inputs from the upperstream layer promotes the model's robustness for test data by merging biases and nonlinear activation functions inside the machine learning model.
This trend can also be observed from the time series of estimated coefficients as shown in figure~\ref{fig:flatplate_result}$(c)$.
The result of case 4 (scalar inputs at downstream layer) shows slight disagreement with the reference data especially for the flow at $(Re_c,\alpha)=(1000,70^\circ)$.

We also investigate the influence \fg{of} the amount of training snapshot 
\fg{in this problem.}
We here consider six numbers of training snapshots $n_{\rm snapshot}=\{50,100,200,300,500,1000\}$.
The averaged $L_2$ error norms of $C_D$ and $C_L$ values for a flat plat wake with $\{Re_c, \alpha\}=\{1000, 70^\circ\}$ are shown in figure~\ref{fig:flat-plate_nos}.
We also present the error variance over the cases 1 to 4 for each number of snapshots as the gray line.
As the number of training snapshots increases, \fg{both the overall error and the variance tend to decrease in all cases.}
{\color{red}Although we can generally see the better estimation by adding scalar inputs at the earlier layers, there are also some cases where this argument looks invalid (e.g., cases 1 and 2 at $n_{\rm snapshot}=300$).
However, we should note that the difference in the $L_2$ error between such cases is $\mathcal{O}(10^{-2})$, and these cases already exhibit acceptable accuracy.
In addition, as we will discuss in figure~\ref{fig:twocy_noise}, the trend that the earlier input leads to the better accuracy can be 
stated more clearly when the problem setting is not too simple.}

Next, we consider a flow over two side-be-side cylinders to investigate the influence on the scalar input placements.
As mentioned above, we utilize the diameter ratio $r$ between cylinders I and II and the distance between two cylinders $g$ as the additional inputs such that three cases for $r=\{1.00, 1.15, 1.30\}$, and six cases for $g=\{0.5,0.7,1.0,1.5,2.0,2.5\}$.
In this example, a single model trained with 18 cases estimates drag and lift coefficients for both cylinders ${\bm y}=(C_{D,1},C_{D,2},C_{L,1},C_{D,2})\in {\mathbb R}^4$ from the vorticity field $\omega$.

\begin{figure*}
    \centering
	\includegraphics[width=\textwidth]{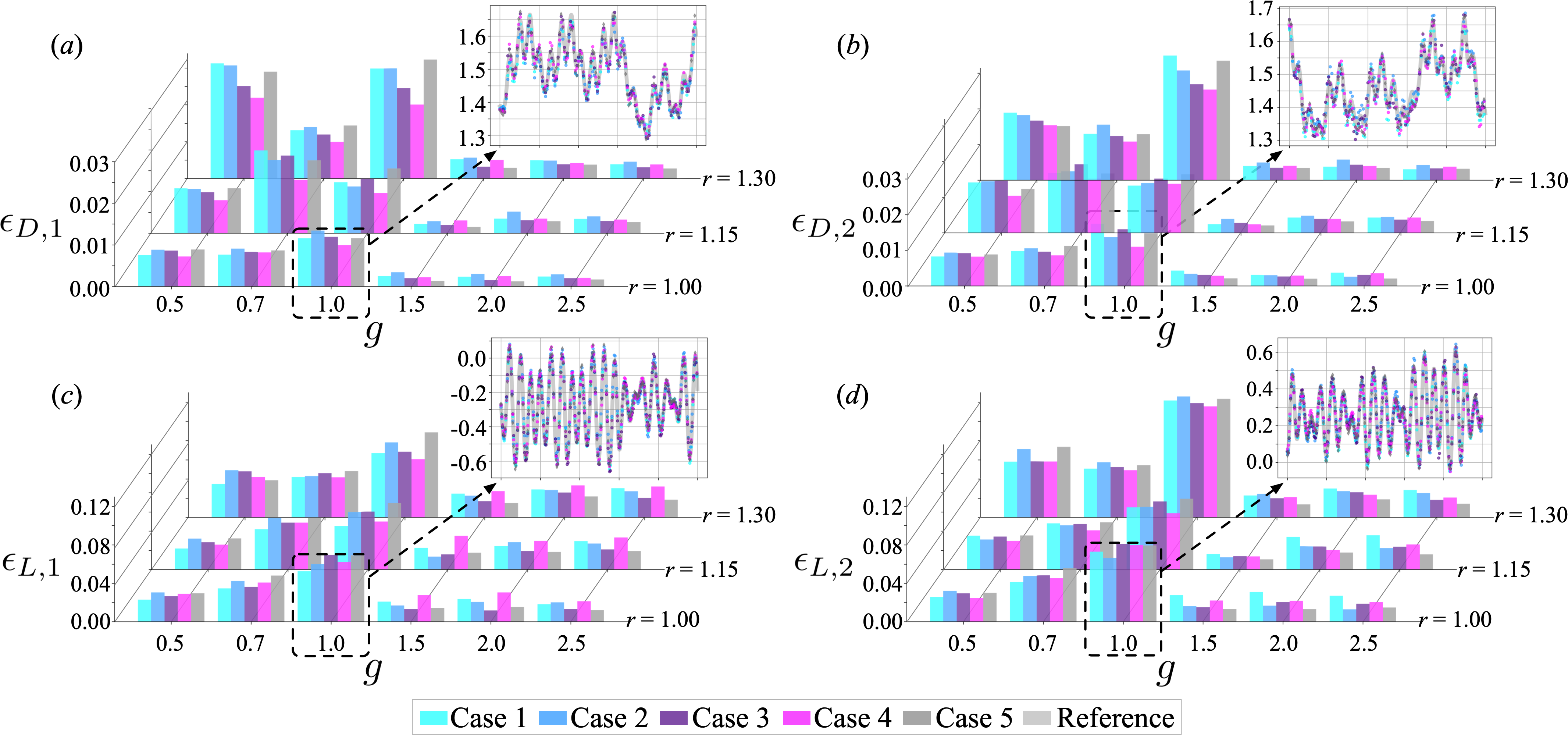}
	\caption{
	Dependence of $C_D$ and $C_L$ estimations on input scalar placements for a flow over two side-by-side cylinders.
	Comparison of the $L_2$ error norm among flow types for each coefficient; $(a)$ $C_D$ of cylinder I, $(b)$ $C_D$ of cylinder II, $(c)$ $C_L$ of cylinder I, and $(d)$ $C_L$ of cylinder II.
	Case 5 denotes the model without the additional scalar value input.
	Time traces of estimated coefficients with $g=1.0$ for each coefficient are also shown.}
	\label{fig:twocy_result}
\end{figure*}
\begin{figure*}
    \centering
	\includegraphics[width=1\textwidth]{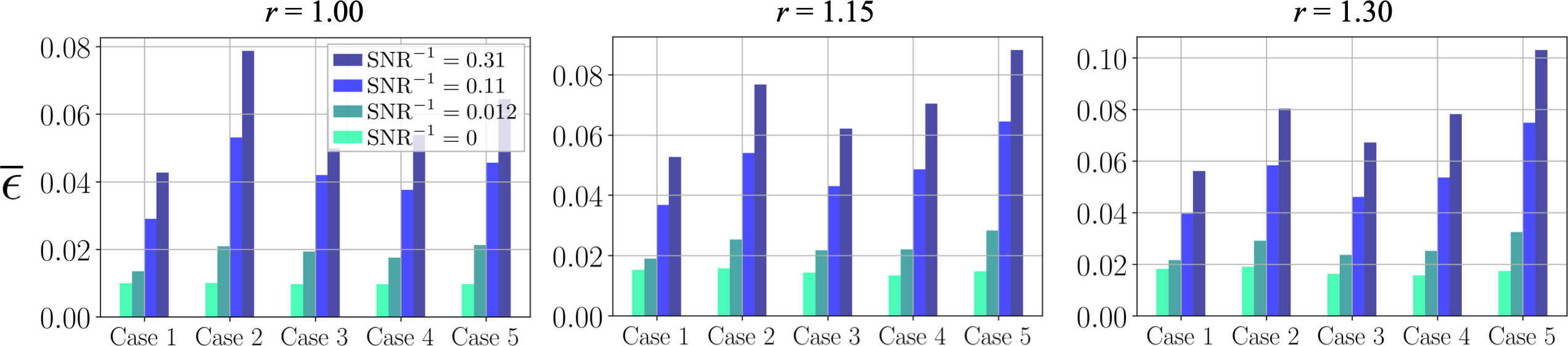}
	\caption{
	Robustness against noisy input for a flow over two side-by-side cylinders.
	The error $\overline{\epsilon}$ represents the averaged $L_2$ error norm over $C_D$ and $C_L$ of both two cylinders for all studied values of distance factor $g$.}
	\label{fig:twocy_noise}
\end{figure*}

The $L_2$ error norms of the estimated four coefficients, $C_D$ and $C_L$ for both cylinders, $\{\epsilon_{D,1},\epsilon_{D,2},\epsilon_{L,1},\epsilon_{L,2}\}$ are summarized in figure~\ref{fig:twocy_result}.
The error rate becomes higher for the flows at $g$ lower than $1.0$, corresponding to chaotic flows, contrary to the flows at higher $g$, i.e., periodic and quasi-periodic flows.
In contrast to the example of flat plate wake, the $L_2$ error norms of $C_L$ tend to be higher than that of $C_D$.
This is likely because of the difference in magnitude --- the value of $C_L$ has the order of ${\cal O}(10^0)$ while that of $C_D$ is ${\cal O}(10^{-1})$.
More concretely, a tiny error, e.g., ${\cal O}(10^{-2})$, cannot be assessed fairly for $C_D$ and $C_L$ since its relative magnitude for them should be different from each other. 
To avoid this issue, standardization or normalization of the data can be a good candidate~\cite{shanker1996effect}.
Noteworthy here is that the dependence of the CNN performance on the input placements of scalar values is lower than that with the flat plate examples.
This suggests that users should care the input placement depending on the data that they handle.

The robustness against noisy input and its dependence on input placements are also examined with the two side-by-side cylinders example, as shown in figure~\ref{fig:twocy_noise}.
A Gaussian noise is added to the input vorticity field.
We consider three magnitudes of the noise whose signal to noise ratio (SNR) is set to ${\rm SNR}^{-1}=\{0.012,0.11,0.31\}$, where ${\rm SNR}=\sigma^2_{\rm Data}/\sigma^2_{\rm Noise}$.
In figure \ref{fig:twocy_noise}, the bars indicate the $L_2$ error norm averaged over $C_D$ and $C_L$ of both two cylinders for all studied values of distance factor $g$ {\color{red}(the result for each configuration is shown independently in Appendix B)}.
As mentioned above, there is no clear difference among cases without noise, presented as the bright-green colorbars.
On the other hand, the cases with the strong magnitude of noise, i.e., blue colored bars, show notable differences over the input placements.
Case 1 (i.e., inputs the scalars at the 1st convolutional layer) reports the least $L_2$ error norm, which is the same trend as that in the flat plate wake example.
This also supports our observation above that the scalar inputs from the upperstream layer enables the model to be robust by merging biases and nonlinear activation functions inside models.
Summarizing above, care should be taken in determining the scalar input placement by considering not only the accuracy for the clean data but also the robustness against noise.

\subsection{Example 2: investigation of CNN parameters for fluid flow analyses with autoencoder}

In this section, let us investigate the influence on CNN parameters for fluid flow analyses using CNN-AE.
Here are the parameters we focus in this study;
\begin{enumerate}
    \item Size of the filter for the convolutional operation (section \ref{sec:sizefilter}).
    \item Compression rate of CNN-AE (section \ref{sec:comprate}).
    \item Padding operation for the convolutional operation (section \ref{sec:padding}).
    \item Input placements of supplemental scalar value (section \ref{sec:2Dplace}).
    \item Methods to reduce/expand the dimension of the data (section \ref{sec:dimredexp}).
    \begin{enumerate}
        \item Up-sampling vs transposed convolution (dimension expansion)
        \item Max vs average pooling (dimension reduction)
    \end{enumerate}
\end{enumerate}

Hereafter, we use a vorticity field of two-dimensional decaying homogeneous isotropic turbulence as the input and output attribute of the AE.
Since it contains various scales of structures and also the data \fg{are} generated with the biperiodical boundary condition, it can be regarded as a suitable example for us to establish a benchmark investigation of CNN parameters, such as filter size, padding operation, and \fg{so on.}

\subsubsection{Size of the filter for the convolutional operation}
\label{sec:sizefilter}

One of the \fg{most advantageous} features of CNN is the filter operation which can extract key spatial structures of the input data to establish a relationship between \fg{the} input and output data.
Because the size of the filter directly relates to the number of weights inside CNN, we can easily expect that an appropriate choice for the filter size can lead to an improvement in estimation ability. 
\fg{This feature is advantageous also in} fluid mechanics applications --- in fact, Fukami et al.~\cite{FFT2019a} utilized a customized CNN, which contains multiple sizes of filters, for fluid flow super-resolution analysis and reported its great ability in handling turbulence compared to a CNN with a single type of filter size.
Otherwise, Lee and You~\cite{LY2019} also capitalized on a CNN which includes multiple sizes of filters for the construction of surrogate modeling for high-fidelity simulation.
Despite that the size of the filter plays a crucial role in CNN-based modeling, the dependence on the filter size for fluid flow data has not been investigated clearly yet.
Here, we consider five sizes of filter $H=\{3,5,7,9,11\}$ and investigate its influence on the ability of CNN-AE.
For this investigation, the training data \fg{are} generated at $Re_0=80.4$, and the size of the latent vector is set to be 1024.

\begin{figure*}
    \centering
		\includegraphics[width=\textwidth]{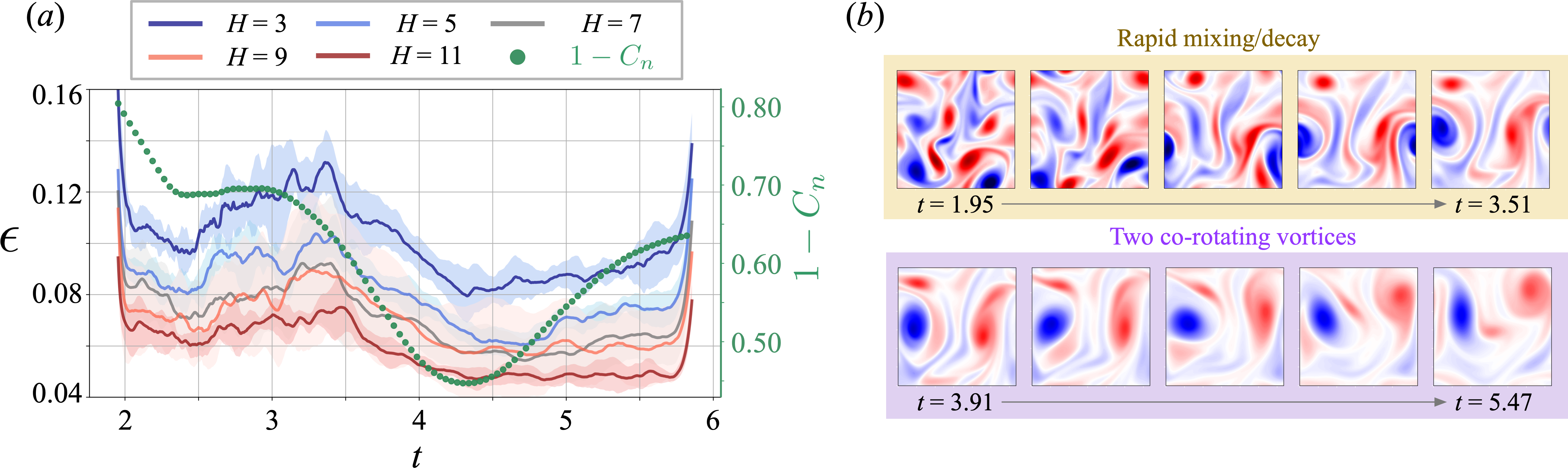}
		\caption{
		{\color{red}$(a)$ Dependence of the $L_2$ error norm on the size of the filter size of CNN-AE.
		Time trace of the $L_2$ error norm and the averaged cosine similarity between each snapshot and the other snapshots are shown.
		The shaded region represents the standard deviation over a three-fold cross-validation.
		$(b)$ Representative flow fields of handled dataset is shown.
		While a rapid mixing and decaying process can be observed for the flows at $1.95<t<3.51$, less spatial variation can be seen on the latter part of the decaying nature.}}
		\label{fig:2Dturb}
\end{figure*}

The time series of $L_2$ error norm of the reconstructed field is shown in figure~\ref{fig:2Dturb}$(a)$.
The model with the larger size of the filters shows lower error level over the time.
This is simply because the number of weight increases for the larger size of the filters and ease the problem setting for the machine learning model.
{\color{red}Along with the time series of the $L_2$ error norms, we also present the averaged cosine similarity in order to seek for the relationship between the variance over the training snapshots and the reconstruction error of the autoencoder.
We define the averaged cosine similarity for each snapshot $C_n$ as
\begin{equation}
    C_n=\frac{1}{N}\sum_{m}\frac{{\bm \omega}_{n}\cdot{\bm \omega}_{m}}{||{\bm \omega}_{n}||~||{\bm \omega}_{m}||},
\end{equation}
where ${\bm \omega}_n$ represents the vorticity field at a certain snapshot and ${\bm \omega}_m$ represents the vorticity field to be compared with ${\bm \omega}_n$, respectively.
The total number of snapshots is denoted as $N$. 
This measure enables us to investigate the influence of the structural and statistical similarities of flow fields over training data on the low-dimensionalization performance.
Note that we present $1-C_n$ so that the trend looks similar to that of the error; $1-C_n=0$ means that the snapshot $n$ is completely the same as the other snapshots, $1-C_n=1$ indicates the orthogonality, and $1-C_n=2$ refers to the same structure with the opposite sign.
As shown in figure~\ref{fig:2Dturb}$(a)$, the time trace of the error basically matches with that of the cosine similarity of each snapshot $1-C_n$.
This result suggests that the autoencoder shows the better performance for the case when the structure is similar to those in the other snapshots.
For example, as shown in figure~12$(b)$, a rapid mixing and decaying process can be seen in the flow fields at $t<3.51$.
Since each snapshot in this period is relatively `unique' among the handled data, i.e., less averaged similarity to the other snapshots, the autoencoder shows less ability in low-dimensionalizing the flow fields.
In contrast, since the flow fields after $t=3.91$ basically contain common structures of two co-rotating vortices, the averaged similarity among the data is relatively high, which leads to the better performance of the autoencoder.
Hence, the training data arrangement and its sampling process affect the error in this particular example associated with the decaying nature.
Moreover, including more snapshots around the beginning and the end of the time series may help to reduce the error for a time period where snapshots tend to be less similar among training data~\cite{nakamura2020extension}.}

\subsubsection{Compression rate of CNN-AE}
\label{sec:comprate}

\begin{figure*}
    \centering
	\includegraphics[width=0.37\textwidth]{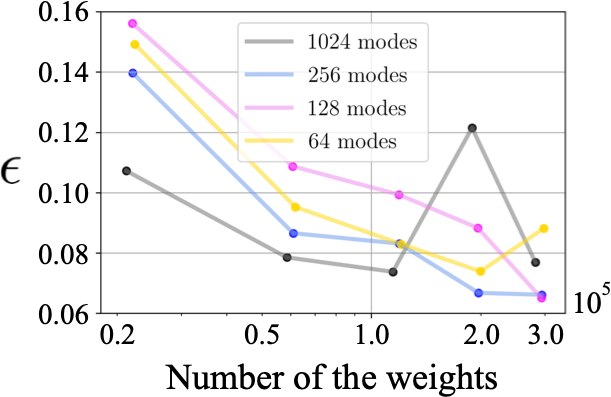}
	\caption{Dependence of the $L_2$ error norm on the number of latent modes and the number of weights for CNN-AE with a two-dimensional decaying turbulence.}
	\label{fig:param}
\end{figure*}

We then investigate the dependence of the compression rate of the CNN-AE on the reconstruction ability.
Generally, the reconstruction performance of AE becomes lower as the number of latent modes decreases~\cite{FHNMF2020}.
Since this is a well-known fact and intuitive for us, we also assess this point with regard to the number of weights inside the CNN-AE, in addition to the number of latent modes, as shown in figure~\ref{fig:param}.
We here present the relationship between the $L_2$ error norm of the reconstructed flow and the number of weights among four cases of latent modes $n_{r}=\{64, 128, 256, 1024\}$.
Since we can increase or decrease various parameters inside CNN including the size of the filter and the number of the layers, we can have the variation for the number of weights over the same numbers of latent modes.
Here, the number of weights is adjusted by changing only the size of the filters $H$.
For the models with the number of weights less than $10^5$, the error decreases as the number of weights increases, as expected.
However, the trend varies for the models with the number of weights larger than $2\times10^5$.
This is likely caused by the fact that an optimization process for weights inside a neural network becomes unstable when the number of weights is massive, which is well known as ``curse of dimensionality."~\cite{Domingos2012}
This result encourages users to be careful in setting the number of weights to avoid the unstable learning process, in addition to the number of latent modes.

\subsubsection{Padding operation for the convolutional operation}
\label{sec:padding}

\begin{figure*}
    \centering
	\includegraphics[width=0.90\textwidth]{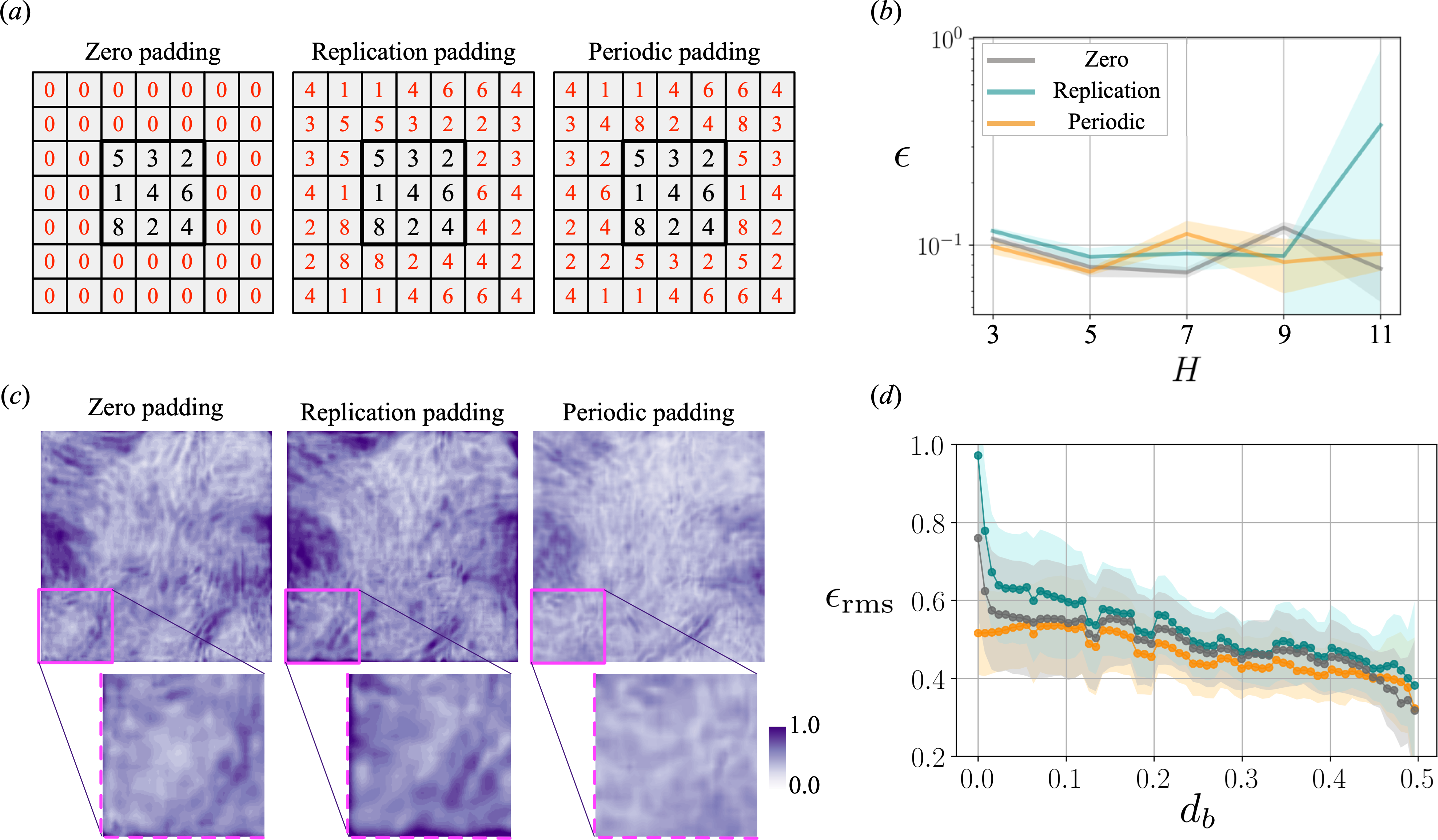}
	\caption{
	$(a)$ Padding operation covered in this study.
	$(b)$ Relationship between the padding operations and the $L_2$ error norm for various filter sizes. The shaded region here represents the standard deviation over a three-fold cross-validation.
	$(c)$ The distribution of root-mean-squared $L_2$ error norm of each padding operation with its filter size of $H=3$. 
	$(d)$ Relationship between the distribution of $\epsilon_{\rm rms}$ and distance from boundaries $d_b$. The ensemble-averaged
	value is shown as a plot while the shaded region represents the standard deviation of the error distribution at each $d_b$.}
	\label{fig:padding}
\end{figure*}

As introduced above, we have recently been able to see various studies of CNN and fluid flows.
\fg{In those studies, however,} \fg{we often see} a suspicious setting for fluid flow analyses --- a zero padding operation at convolutional layers.
Fundamentally, the convolutional operation usually reduces the data size $L_w\times L_h$ to $(L_w-2\lfloor H/2\rfloor)\times(L_h-2\lfloor H/2\rfloor)$\fg{, but} we often do not want the data size to be reduced via convolutional operations because the dimension reduction using pooling layers is known as the better way to acquire the robustness for spatial sensitivity and noise~\cite{He2016}.
To avoid the dimensional reduction at the convolutional layers, it is very common to add zero values around the data (called {\it zero padding}), as shown in figure~\ref{fig:padding}$(a)$.
By giving additional values around an image, the dimension of the data can be retained through the convolutional operation.
Although \fg{this zero padding} is \fg{a commonly} used technique, it is questionable from the perspective of boundary conditions, especially with numerically generated data.
For instance, it can be easily suspected whether we can use the zero padding for the present two-dimensional decaying turbulence having the spatial biperiodic condition, or not.
Here, we compare three types of padding operations, 1. zero padding, 2. replication padding, and 3. periodic padding.
As illustrated in figure~\ref{fig:padding}$(a)$, the replication padding is a mirroring operation which embeds symmetrical value regarding the boundary of the data.
This operation is also non-physical \fg{similarly to} zero padding, although the boundary \fg{values} become smooth.
The other covered operation is the periodic padding, which faithfully follows the boundary condition of the present turbulence data.

We investigate these three operations with various sizes of the filter $H$ from $3$ to $11$, as presented in figure~\ref{fig:padding}$(b)$.
\fg{No} clear difference is observed up to $H\leq9$, while the error variance becomes significantly unstable using the replication padding with $H=11$ as can be seen from its standard deviation shown with a shaded region.
This negative influence of replication padding with the larger size of filters can be regarded as natural because more nonphysical values are given, which suggests that these values would affect the convolutional operation inside \fg{the} CNN.
In contrast, what is striking here is that \fg{the} models with zero padding show the same error level as \fg{those} with periodic padding.
This is likely because the influence of inserting zero values near the boundaries is much less than that with nonphysical values by the replication padding 
in \fg{the operation of} equation~\ref{eq:CNN}.

Let us also examine the local errors in the region near the boundaries.
The distribution of root-mean-squared local $L_2$ error norm $\epsilon_{\rm rms}$ with each padding operation is shown in figure~\ref{fig:padding}$(c)$.
We use the filter size of $H=3$, as an example.
While the error magnitude of replication padding is relatively larger at the region near the boundaries (shown as pink dotted lines), we can avoid the boundary influence by utilizing the zero padding and the periodic padding.

We also assess the dependence of the local $\epsilon_{\rm rms}$ on the distance from the nearest boundary $d_b$.
Since the present domain is a square box with $L_x=L_y=1$, the distance $d_b$ can simply be defined as
\begin{equation}
d_b=\min(d_x, d_y) ,
\end{equation}
where
\begin{equation}
d_x=\min(x,1-x),\ 
d_y=\min(y,1-y) .
\end{equation}
The relationship between $d_b$ and $\epsilon_{\rm rms}$ is presented in figure~\ref{fig:padding}$(d)$.
In the figure, the ensemble average over the points having the same $d_b$ is shown as a plot while the shaded region represents the standard deviation.
In the proximity of the boundaries, i.e., $d_b \simeq 0$, the error rate jumps to $\epsilon_{\rm rms} \simeq 1.0$ for the case of the replication padding, which is the largest among the covered cases.
The error rate of the zero padding also increases at $d_b\simeq 0$; however, the error rapidly (at $d_b>0.05$) decreases to the almost same level as that with periodic padding.
Concluding the discussions above, the optimal choice of padding operation may differ among flows with different boundary conditions.

\subsubsection{Input placements of supplemental scalar value}
\label{sec:2Dplace}

As introduced above, the supplemental scalar values are often utilized to improve the estimation ability of CNN for fluid flow analyses.
Similar to the investigation for the influence on input placements of scalar values with the CNN-MLP model in section~\ref{sec:cnn-mlp}, let us also examine this point with the CNN-AE for the decaying isotropic homogeneous turbulence.
As briefly explained in section~\ref{sec:covered_cnn}, we consider four cases of input placements for additional scalar values, i.e, the 1st convolutional layer (case 1), the 4th convolutional layer (case 2), the latent space (case 3), and the 16th convolutional layer (case 4).
As illustrated in figure~\ref{fig:CNN-AE}, the scalar values are first expanded with MLP and then reshaped into two-dimensional sectional data to concatenate with outputs provided at the hidden layer inside the CNN-AE.
As supplemental scalar values, we utilize two types of the Reynolds number, the initial Reynolds number $Re_0$ and the instantaneous Taylor Reynolds number $Re_{\lambda}$.
While the initial Reynolds number $Re_0$ represents the initial condition and governs the decaying nature of flows, the instantaneous Taylor Reynolds number $Re_{\lambda}$ contains information of an instantaneous flow snapshot.
In addition, we also consider the case with no additional scalar values as case 5.

\begin{figure*}
    \centering
	\includegraphics[width=\textwidth]{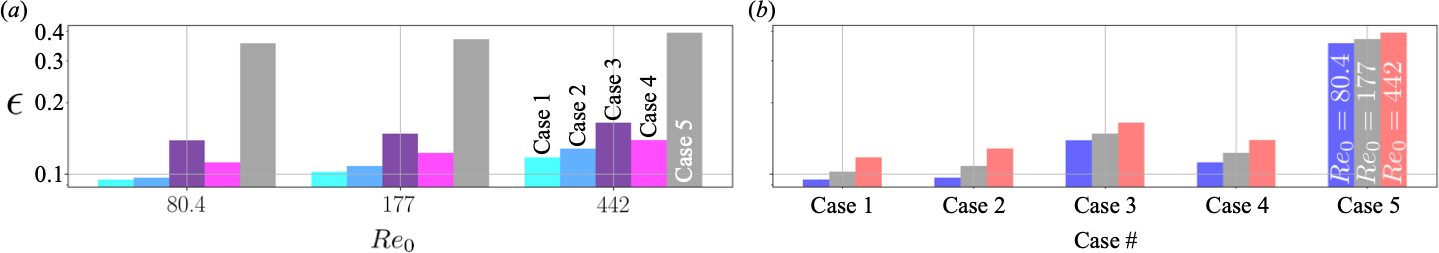}
	\caption{
	Dependence of supplemental scalar value $\{Re_0,Re_\lambda\}$ on input placements for the CNN-AE with two-dimensional decaying turbulence.
	$(a)$ Relationship between the $L_2$ error norm and initial Reynolds number depending on input placement.
	$(b)$ Relationship between the $L_2$ error norm and input placement depending on initial Reynolds number.}
		\label{fig:re_dep}
\end{figure*}

The dependence of the $L_2$ error norm of reconstructed flows on the input scalar placements is summarized in figure~\ref{fig:re_dep}.
As we can clearly see, case 5 (without the scalar inputs) reports higher error than other cases.
This implies that the use of supplemental scalars \fg{is beneficial also} for low-dimensionalization.
For the cases with scalar inputs, case 1 (input at the first layer) shows the best estimation among the covered cases, which is analogous to the investigation with the CNN-MLP model above.
On the other hand, case 3 (input from the latent space) reports the worst estimation.
This is likely due to the difference in the number of weights given at the MLP part for merging the Reynolds number input and the main part of CNN-AE.
Summarizing above, the results indicate that scalar inputs at upstream layers helps the estimation for CNN-AE, as well as the CNN-MLP model, and additionally a sufficient number of weights are required before merging the input scalars with CNN-AE.

\subsubsection{Dimensional reducing/expanding methods}
\label{sec:dimredexp}

\begin{figure*}
    \centering
	\includegraphics[width=0.85\textwidth]{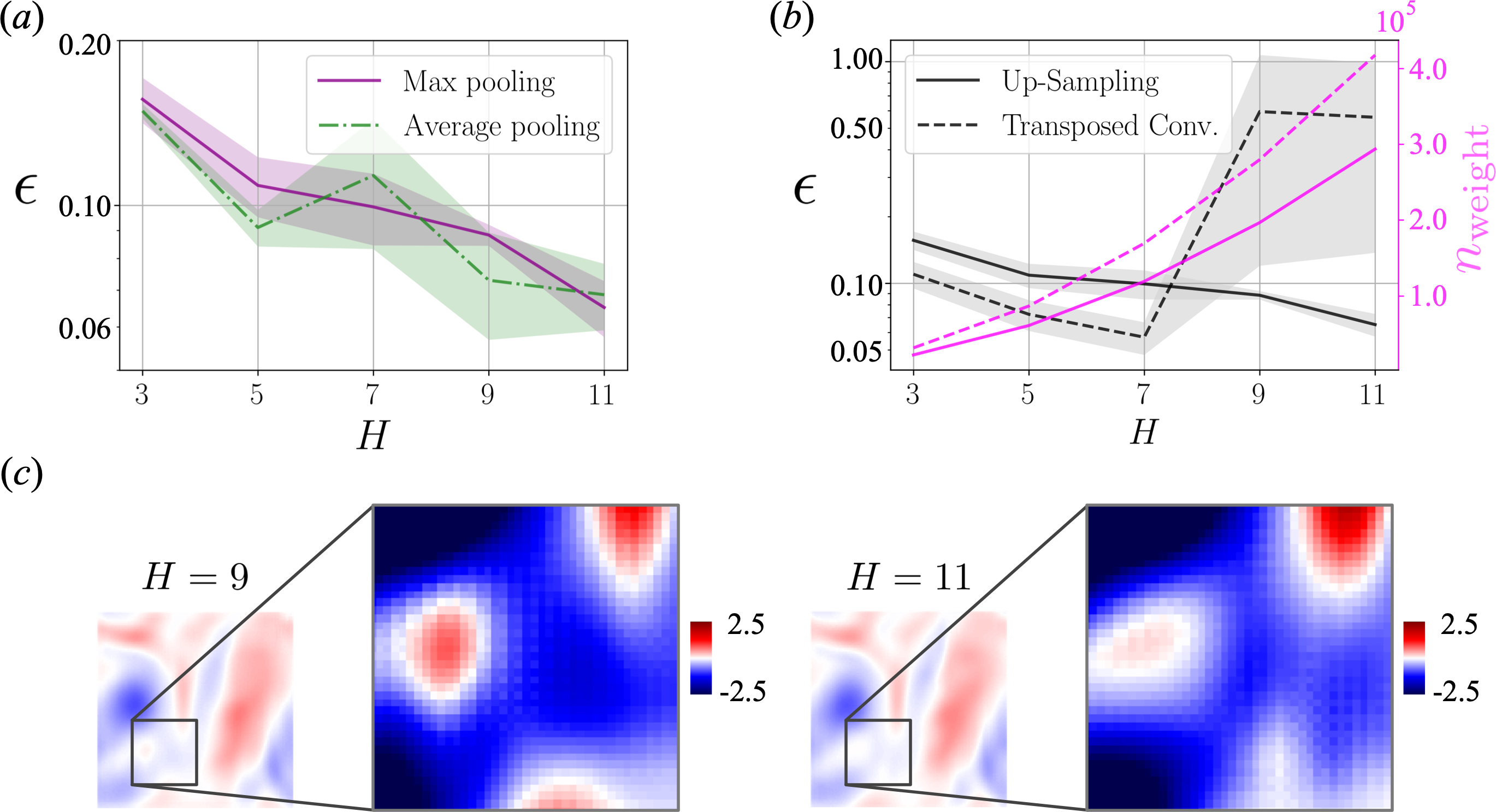}
	\caption{
	Relationship between the $L_2$ error norm $\epsilon$ and filter size $H$ for $(a)$ dimension reduction methods (max and average pooling), and $(b)$ dimension expansion methods (up-sampling and transposed convolution).
	The shaded region represents the standard deviation over a three-fold cross-validation.
	$(c)$ The estimated fields of models using transposed convolution with the filter size of $H=9$ and $11$ are shown.
	A checkerboard artifact can be seen due to the unsuccessful training by using the transposed convolution with the large filter size.}
	\label{fig:pool-up}
\end{figure*}

Finally, we investigate the effect of the dimensional reduction and expansion methods for CNN performance.
The relationship between the $L_2$ error norm and the dimension reduction ways are summarized in figure~\ref{fig:pool-up}$(a)$.
We compare the max and average poolings, as introduced in section~\ref{sec:cnn}.
As an example, the number of latent variables is set to be 128 with five cases of filter size $H=\{3,5,7,9, 11\}$.
As shown, the error with both methods are approximately at the same level for every case.
This result indicates that the present AE is not very sensitive to the pooling operations.

We also check the influence of dimensional expansion methods, as presented in figure~\ref{fig:pool-up}$(b)$.
As introduced in section~\ref{sec:covered_cnn} and figure~\ref{fig:CNN_operation}, the significant difference between the transposed convolution and the up-sampling is whether there are trainable weights (filter) or not.
Therefore, the number of weights inside a model $n_{\rm weight}$ increases when up-sampling layers are replaced with transposed convolutional layers, as shown by the pink curve in figure \ref{fig:pool-up}$(b)$.

As summarized here, the $L_2$ error norm of the models with up-sampling operation gradually decreases as the filter size increases.
This exactly corresponds to the increase of the number of weights.
In contrast, the variance of the model with the transposed convolutional operation is significantly unstable with $H\geq9$ where the number of the weights reaches approximately $3.0\times10^5$.
This is analogous to the observation in section~\ref{sec:comprate}, known as ``curse of dimensionality."

The unsuccessful training procedure of the network with transposed convolution can also be found with the appearance of checkerboard artifacts~\cite{odena2016deconvolution}.
The checkerboard artifacts tend to appear with strided-transposed convolution, especially when the network fails to update the weights properly through its training procedure.
In our cases, the fields estimated by the networks with the filter size of $H=9$ and 11, which reported significantly worse $L_2$ error norm in figure~\ref{fig:pool-up}$(b)$, exhibit such artifacts, as shown in figure \ref{fig:pool-up}$(c)$.
This is again due to the excessive number of weights contained in the network and a sign that the transposed convolutional operations with a relatively large filter size negatively affect the training procedure.
We note that the present model with the transposed convolution generally shows better performance with less amount of weights, i.e., $H\leq7$.
Hence, the choice for the dimension expansion and reduction methods inside CNNs should be cared depending on considered flows and their model configurations.

\section{Conclusions}
\label{sec:conclusion}
\begin{figure*}
    \centering
	\includegraphics[width=0.8\textwidth]{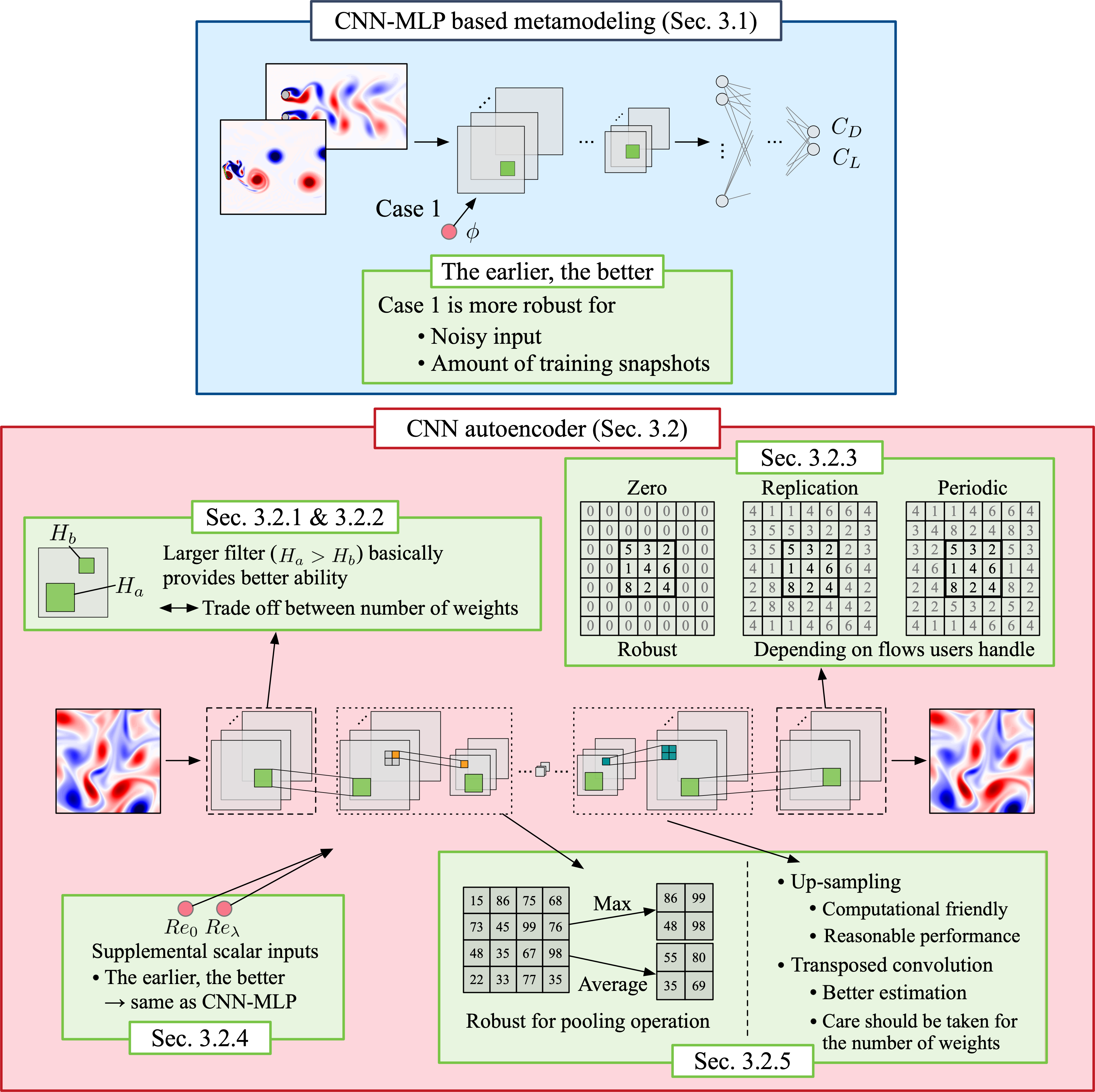}
	\caption{Graphical conclusion of the present study. }
	\label{fig:conc}
\end{figure*}

We investigated the applicability of convolutional neural networks (CNNs), which have been utilized as a powerful tool in scientific machine learning, for fluid flow analyses.
Especially, we focused on the CNN application with additional scalar input information such that ${\bm y}={\cal F}({\bm x},{\phi})$.
Capitalizing on the canonical fluid flow data with the perspectives on metamodeling for aerodynamics characteristics and low-dimensionalization, we attempted to clear vague portions of CNN and fluid flow analyses as graphically summarized in figure~\ref{fig:conc}.

We first considered the CNN-MLP model with additional scalar inputs, by considering flows around \fg{an inclined} flat plate and two side-by-side cylinders.
The model attempted to estimate drag and lift coefficients from a vorticity field.
The supplemental scalar values were added to the model at four different placements and we investigated the influence on the input placements for the scalar values.
{\color{red}Although the response of the machine-learned model depends on target flows and associated problem settings,} we observed {\color{red}a general trend that} better estimations {\color{red}can be attained by placing the} scalar inputs at earlier layers\fg{,} not only in terms of accuracy but also robustness, especially under relatively tough problem settings, i.e., less amount of training data and noisy input.
This is likely because a neural network \fg{gains robustness against} the test data by merging supplemental information with biases and nonlinear activation functions inside the neural network.
Although our investigation was performed using the same flow configuration as that used for training, it is widely known that neural networks can acquire the robustness against unseen flows by preparing a proper training data set~\cite{MFZF2020,HFMF2020a,HFMF2020b}.
We believe that our present findings can also be useful by unifying these previous studies related to the applicability to unseen flow data.

We also examined the influence on CNN parameters considering an autoencoder (AE)-based low-dimensionalization.
Through the investigations for the filter size and the compression rate dependence, we found that the larger number of weights contained in the model can lead to the better estimation.
However, care should be taken for cases which contains too many weights since it causes unstableness of the training procedure of CNN.
{\color{red}The dependence on the padding operation for convolution was also examined in a quantitative manner.
Although the periodic padding shows the best performance since the present decaying turbulence is simulated under the bi-periodic condition, the zero padding is also found to give a reasonable overall accuracy.
Users can choose the optimal padding operation depending on the boundary conditions of flow data.}
Similar to the investigation on the CNN-MLP models, we also checked the influence of input placements for supplemental scalar values on the estimation ability of CNN-AE.
Here, we also observed that the scalar input at earlier layers leads to improve the estimation more compared to the other cases.
Finally, the choice of dimensional reduction/expansion methods was also considered.
The machine learning model is \fg{less sensitive} the dimensional reduction methods, while it is sensitive to the expansion methods, i.e., up-sampling and transposed convolution.
Since the model with the transposed convolution contains large amount of weights than that with up-sampling layers, the model may \fg{fall} into the ``curse of dimensionality."
Based on the above, users can attempt to choose an appropriate combination of weights depending on their problem setting and computational environment.

The present investigations enable us to notice some remaining issues and considerable extensions of CNN and fluid flow analyses.
One of them is applications of CNN to unstructured mesh fluid flow data.
As expressed in section~\ref{sec:cnn}, the basic principle inside CNN is taking filters for fluid flow {\it image} handled on structured/discretized grids arranged with a {\it uniform} manner.
However, as readers have already noticed, we often encounter unstructured data for \fg{many} fluid flow analyses, e.g., flows around an airfoil at practical Reynolds numbers. 
Unfortunately, the conventional CNN used through this study cannot be applied directly to these data on unstructured mesh.
To overcome this issue, several ideas have recently been proposed, e.g., graph convolutional neural networks~\cite{ogoke2020graph,tencer2020enabling}, PointNet~\cite{kashefi2020point}, PhyGeoNet~\cite{GSW2020}, and Voronoi-tessellation aided CNN model~\cite{FukamiVoronoi}.
We can expect that the present knowledge and the ways for analyzing CNN parameters with fluid flows can be extended to the aforementioned models.
Otherwise, readers' interest in a practical manner must arrive at the perspective on the interpretability of results provided by CNNs.
For real-world applications, we often desire grounds of estimation by a model (not only machine learning).
Actually, some studies have tackled this point from various views, e.g., the relationship between machine-learned results and vortical motion~\cite{KL2020}, uncertainty quantification~\cite{MFRFT2020}, and visualization inside CNNs~\cite{MFZF2020}.
{\color{red}In addition, the applicability of a supervised machine-learned model for test situations which are completely different from the training data regime is also one of the key factors towards practical applications~\cite{erichson2020}.}
Although it is still an on-going area of research in fluid mechanics and CNN, these trends may be a good direction toward practical applications.

\section*{Acknowledgements}

Masaki Morimoto, Kai Fukami, and Koji Fukagata were supported by JSPS KAKENHI Grant Number 18H03758, 21H05007.
Masaki Morimoto and Kai Fukami thank Mr. Taichi Nakamura (Keio University) for insightful discussions.
Aditya G. Nair acknowledges the support by UNR VP Research Startup PG19679.

\section*{Data availability statement}
The datasets generated during and/or analysed during the current study are available from the corresponding author on reasonable request.

\section*{Appendix A: Details of neural networks}
Here, we provide \fg{the} detailed information on the neural networks covered in this study.
In tables~\ref{tab:structure_CNN-MLP_fp} and \ref{tab:structure_CNN-MLP_twocy}, we provide the structure of the CNN-MLP model for force coefficient estimation.
The main stream of the neural networks is shown \fg{in} the left half of the tables, whereas the right half shows the MLP-CNN part which expands the supplemental scalar values onto two dimensional data to match the data size with that of the target layer to \fg{be} concatenate\fg{d} with.
The last layer of the MLP-CNN part is then concatenated with the main stream network.

Table~\ref{tab:structure_CNNAE} shows the structure of CNN-AE utilized in section~\ref{sec:2Dplace}.
The supplemental Reynolds numbers are expanded using the MLP-CNN part shown \fg{in} the right half of the table.
They are then concatenated with the mainstream network at four different placements.

\begin{table}
    \centering
    \caption{Network structure of the CNN-MLP model used for force coefficient estimation of flow over a flat plate. The left half of the table shows a mainstream of the network, while the right half shows the part which expands scalar values to fed into the mainstream network. For Conv2D layers, the size of filter $H$ and the number of filters $n$ are denoted as $(H, H, n)$.}
    \label{tab:structure_CNN-MLP_fp}
    \begin{tabular}{lccc|lccc}\hline
         Layer                   & Data size    & Activation & Case & Layer                  & Data size & Activation & Case\\ \hline
         Input (vorticity field) & (148,132,1)  & --         & all        & Input ($Re_c, \alpha$) & (2)       &            & 1,2\\
         Conv2D (7,7,16)         & (148,132,32) & ReLU       & all        & Dense                  & (16)      & ReLU       & 1,2\\
         Concatenate             & (148,132,34) & --         & 1          & Dense                  & (32)      & ReLU       & 1,2\\
         Max Pooling             & (74,66,32)   & --         & all        & Dense                  & (64)      & ReLU       & 1,2\\
         Conv2D (7,7,16)         & (74,66,32)   & ReLU       & all        & Dense                  & (256)     & ReLU       & 1,2\\
         Max Pooling             & (37,33,32)   & --         & all        & Dense                  & (512)     & ReLU       & 1,2\\
         Conv2D (7,7,16)         & (37,33,16)   & ReLU       & all        & Dense                  & (1024)    & ReLU       & 1,2\\
         Concatenate             & (37,33,18)   & --         & 2        & Dense                  & (1221)    & ReLU       & 1,2\\
         Conv2D (7,7,8)          & (37,33,8)    & ReLU       & all        & Reshape                & (37,33,1) & --         & 1,2\\
         Conv2D (7,7,4)          & (37,33,4)    & ReLU       & all        & Conv2D (7,7,2)         & (37,33,2) & ReLU       & 1,2\\
         Conv2D (7,7,1)          & (37,33,1)    & ReLU       & all        & UpSampling             & (74,66,2) & --         & 1\\
         Reshape                 & (1024)       & --         & all        & Conv2D (7,7,2)         & (74,66,2) & ReLU       & 1,2\\
         Concatenate             & (1026)       & --         & 3          & UpSampling             & (148,132,2) & --       & 1\\
         Dense                   & (1024)       & ReLU       & all        & Conv2D (7,7,16)        & (148,132,2) & ReLU     & 1,2\\
         Dense                   & (256)        & ReLU       & all\\
         Dense                   & (64)         & ReLU       & all\\
         Concatenate             & (66)         & --         & 4\\
         Dense                   & (32)         & ReLU       & all\\
         Dense                   & (16)         & ReLU       & all\\
         Dense                   & (2)          & Linear     & all\\ \hline
    \end{tabular}
\end{table}

\begin{table}
    \centering
    \caption{Network structure of the CNN-MLP model used for force coefficient estimation of flow over two side-by-side cylinders. The left half of the table shows a mainstream of the network, while the right half shows the part which expands scalar values to fed into the mainstream network. For Conv2D layers, the size of filter $H$ and the number of filters $n$ are denoted as $(H, H, n)$.}
    \label{tab:structure_CNN-MLP_twocy}
    \begin{tabular}{lccc|lccc}\hline
         Layer                   & Data size    & Activation & Case & Layer          & Data size & Activation & Case\\ \hline
         Input (vorticity field) & (240,448,1)  & --         & all  & Input ($g, r$) & (2)       & --         & 1,2\\
         Conv2D (7,7,32)         & (240,448,32) & ReLU       & all  & Dense          & (16)      & ReLU       & 1,2\\
         Concatenate             & (240,448,34) & --         & 1    & Dense          & (32)      & ReLU       & 1,2\\
         Max Pooling             & (120,224,32) & --         & all  & Dense          & (64)      & ReLU       & 1,2\\
         Conv2D (7,7,32)         & (120,224,32) & ReLU       & all  & Dense          & (256)     & ReLU       & 1,2\\
         Max Pooling             & (60,112,32)  & --         & all  & Dense          & (420)     & ReLU       & 1,2\\
         Concatenate             & (60,112,34)  & --         & 2    & Reshape        & (15,28,1) & --         & 1,2\\
         Conv2D (7,7,32)         & (60,112,32)  & ReLU       & all  & Conv2D (7,7,2) & (15,28,2) & ReLU       & 1,2\\
         Max Pooling             & (30,56,32)   & --         & all  & UpSampling     & (30,56,2) & --         & 1,2\\
         Conv2D (7,7,32)         & (30,56,32)   & ReLU       & all  & Conv2D (7,7,2) & (30,56,2) & ReLU       & 1,2\\
         Conv2D (7,7,16)         & (30,56,16)   & ReLU       & all  & UpSampling     & (60,112,2) & --        & 1,2\\
         Max Pooling             & (15,28,16)   & --         & all  & Conv2D (7,7,2) & (60,112,2) & ReLU      & 1,2\\
         Conv2D (7,7,16)         & (15,28,16)   & ReLU       & all  & UpSampling     & (120,224,2) & --       & 1\\
         Conv2D (7,7,8)          & (15,28,8)    & ReLU       & all  & Conv2D (7,7,2) & (120,224,2) & ReLU     & 1,2\\
         Conv2D (7,7,4)          & (15,28,4)    & ReLU       & all  & UpSampling     & (240,448,2) & --       & 1\\
         Reshape                 & (1680)       & --         & all  & Conv2D (7,7,16) & (240,448,2) & ReLU    & 1,2\\
         Concatenate             & (1682)       & --         & 3\\
         Dense                   & (1024)       & ReLU       & all\\
         Dense                   & (512)        & ReLU       & all\\
         Dense                   & (256)        & ReLU       & all\\
         Dense                   & (128)        & ReLU       & all\\
         Dense                   & (64)         & ReLU       & all\\
         Concatenate             & (66)         & --         & 4\\
         Dense                   & (32)         & ReLU       & all\\
         Dense                   & (16)         & ReLU       & all\\
         Dense                   & (4)          & Linear     & all\\ \hline
    \end{tabular}
\end{table}

\begin{table}
    \centering
    \caption{Network structure of the CNN-AE for two dimensional decaying turbulence. The left half of the table shows a mainstream of the network, while the right half shows the part which expands scalar values to fed into the mainstream network. For Conv2D layers, the size of filter $H$ and the number of filters $n$ are denoted as $(H, H, n)$.}
    \label{tab:structure_CNNAE}
    \begin{tabular}{lccc|lccc}\hline
         Layer                   & Data size    & Activation & Case & Layer          & Data size & Activation & Case\\ \hline
         Input (vorticity field) & (256,256,1)  & --         & all  & Input ($Re_0,Re_\lambda$) & (2) & --    & all\\
         Conv2D (7,7,32)         & (256,256,32) & ReLU       & all  & Dense          & (16)        & ReLU     & all\\
         Concatenate             & (256,56,34)  & --         & 1    & Dense          & (32)        & ReLU     & all\\
         Max Pooling             & (128,128,32) & --         & all  & Dense          & (64)        & ReLU     & all\\
         Conv2D (7,7,32)         & (128,128,32) & ReLU       & all  & Dense          & (256)       & ReLU     & 1,2,4\\
         Max Pooling             & (64,64,32)   & --         & all  & Reshape        & (16,16,1)   & ReLU     & 1,2,4\\
         Conv2D (7,7,32)         & (64,64,32)   & ReLU       & all  & Conv2D (7,7,2) & (16,16,2)   & ReLU     & 1,2,4\\
         Max Pooling             & (32,32,32)   & --         & all  & UpSampling     & (32,32,2)   & --       & 1,2,4\\
         Concatenate             & (32,32,34)   & ReLU       & 2    & Conv2D (7,7,2) & (32,32,2)   & ReLU     & 1,2,4\\
         Conv2D (7,7,32)         & (32,32,32)   & ReLU       & all  & UpSampling     & (64,64,2)   & --       & 1\\
         Max Pooling             & (16,16,32)   & --         & all  & Conv2D (7,7,2) & (64,64,2)   & ReLU     & 1,2,4\\
         Conv2D (7,7,16)         & (16,16,32)   & ReLU       & all  & UpSampling     & (128,128,2) & --       & 1\\
         Max Pooling             & (8,8,32)     & --         & all  & Conv2D (7,7,2) & (128,128,2) & ReLU     & 1,2,4\\
         Conv2D (7,7,32)         & (8,8,32)     & ReLU       & all  & UpSampling     & (255,256,2) & --       & 1\\
         Conv2D (7,7,8)          & (8,8,8)      & ReLU       & all  & Conv2D (7,7,2) & (256,256,2) & ReLU     & 1,2,4\\
         Conv2D (7,7,4)          & (8,8,4)      & ReLU       & all  & Conv2D (7,7,2) & (256,256,2) & ReLU     & 1,2,4\\
         Conv2D (7,7,2)          & (8,8,2)      & ReLU       & all\\
         Concatenate             & (8,8,4)      & ReLU       & 3\\
         Conv2D (7,7,2)          & (8,8,2)      & ReLU       & all  & Dense          & (64)        & ReLU     & 3\\
         Conv2D (7,7,2)          & (8,8,2)      & ReLU       & all  & Reshape        & (8,8,1)   & ReLU       & 3\\
         Conv2D (7,7,4)          & (8,8,4)      & ReLU       & all  & Conv2D (7,7,2) & (8,8,2) & ReLU         & 3\\
         Conv2D (7,7,8)          & (8,8,8)      & ReLU       & all  & Conv2D (7,7,2) & (8,8,2) & ReLU         & 3\\
         Conv2D (7,7,32)         & (8,8,32)     & ReLU       & all\\
         UpSampling              & (16,16,32)   & --         & all\\
         Conv2D (7,7,32)         & (16,16,32)   & ReLU       & all\\
         UpSampling              & (32,32,32)   & --         & all\\
         Concatenate             & (32,32,34)   & ReLU       & 4\\
         Conv2D (7,7,32)         & (32,32,32)   & ReLU       & all\\
         UpSampling              & (64,64,32)   & --         & all\\
         Conv2D (7,7,32)         & (64,64,32)   & ReLU       & all\\
         UpSampling              & (128,128,32) & --         & all\\
         Conv2D (7,7,32)         & (128,128,32) & ReLU       & all\\
         UpSampling              & (256,256,32) & --         & all\\
         Conv2D (7,7,32)         & (256,256,32) & ReLU       & all\\
         Conv2D (7,7,1)          & (256,256,1)  & Linear       & all\\ \hline
    \end{tabular}
\end{table}

\section*{{\color{red}Appendix B: Detailed observation for the robustness of the CNN-MLP model against noisy input}}
\begin{figure*}
    \centering
	\includegraphics[width=\textwidth]{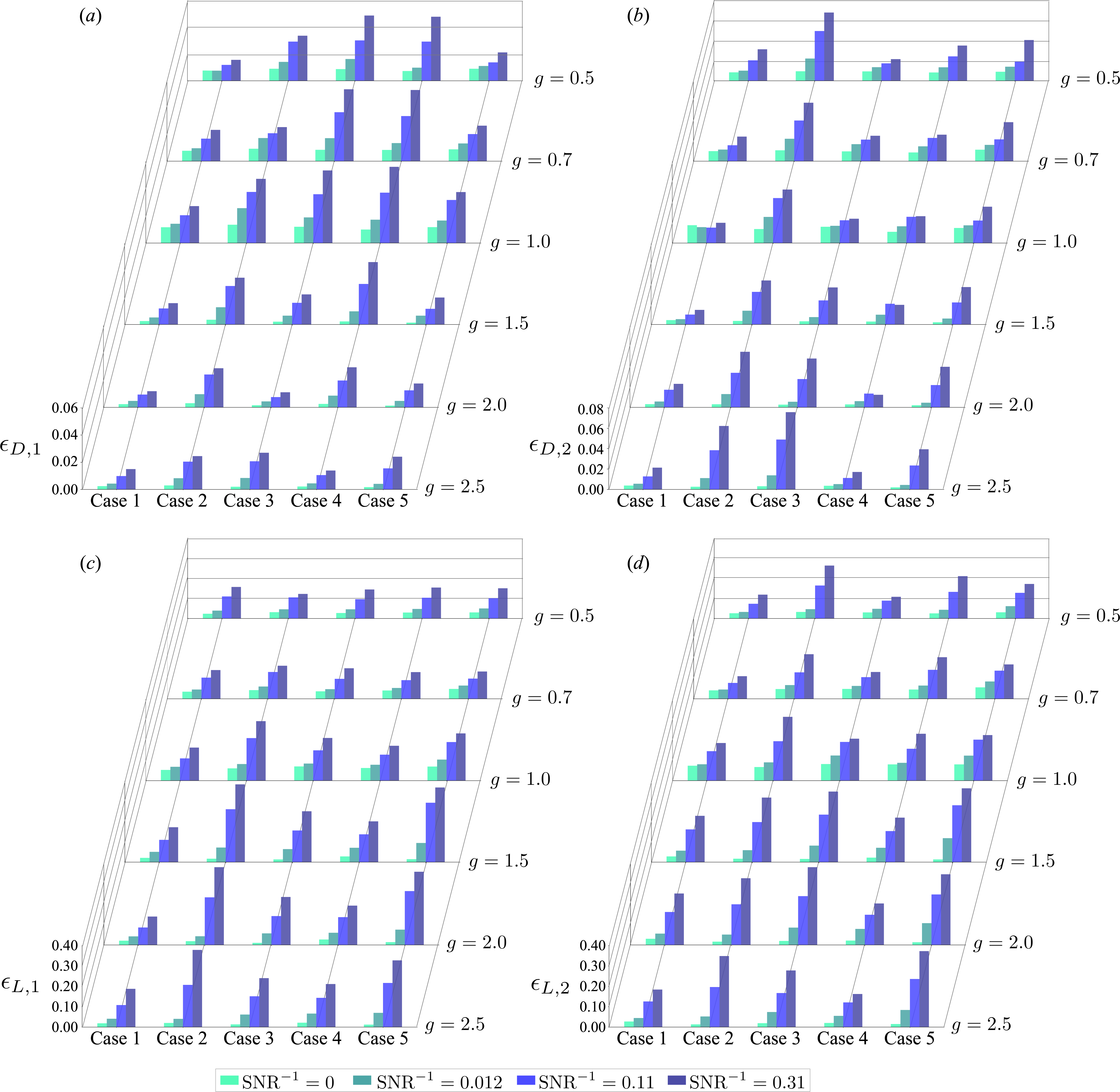}
	\caption{{\color{red}Robustness against noisy input for a flow over two side-by-side cylinders with its radius ratio of $r=1.00$.}}
	\label{fig:A1}
\end{figure*}
\begin{figure*}
    \centering
	\includegraphics[width=\textwidth]{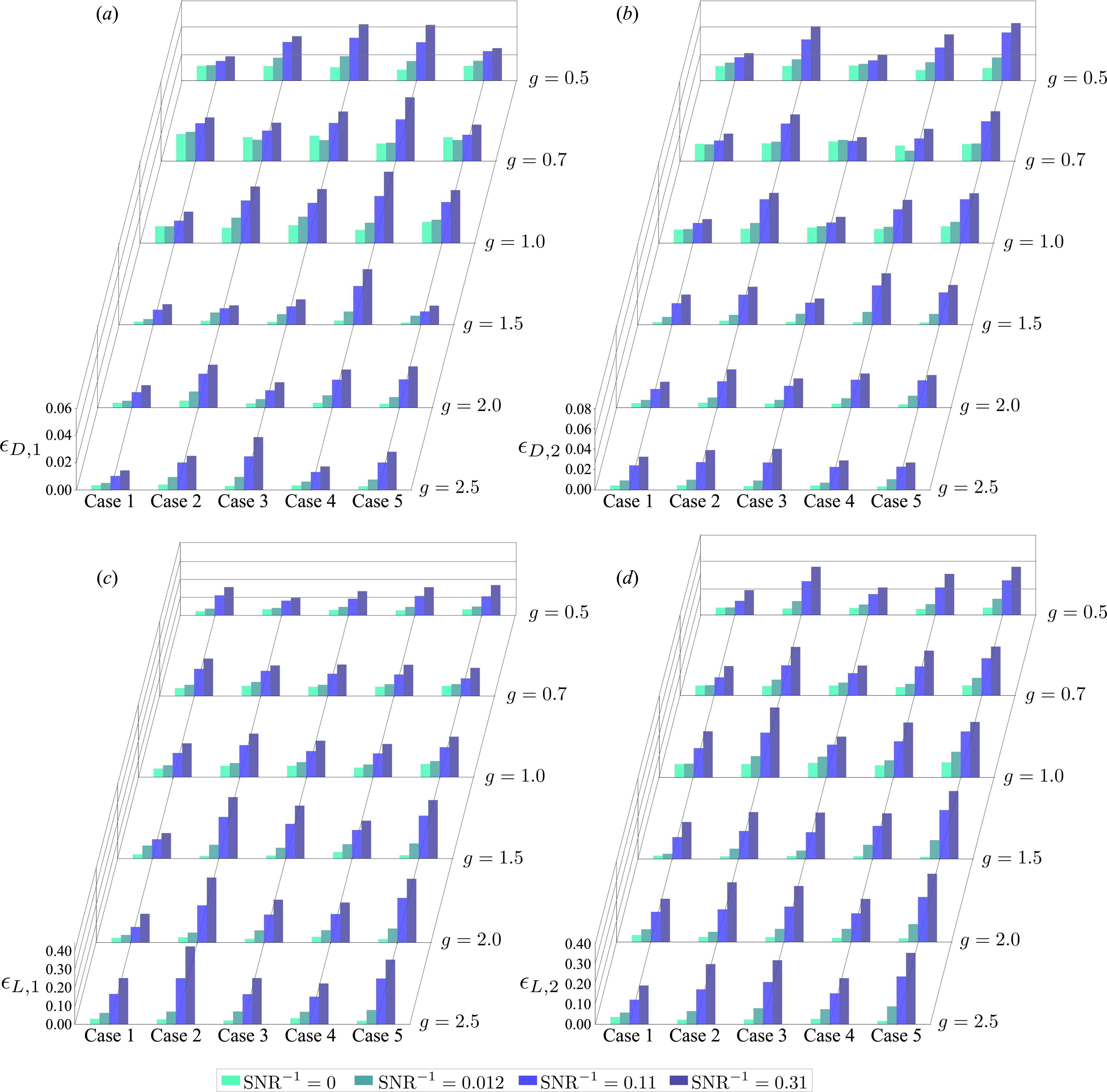}
	\caption{{\color{red}Robustness against noisy input for a flow over two side-by-side cylinders with its radius ratio of $r=1.15$.}}
	\label{fig:A2}
\end{figure*}
\begin{figure*}
    \centering
	\includegraphics[width=\textwidth]{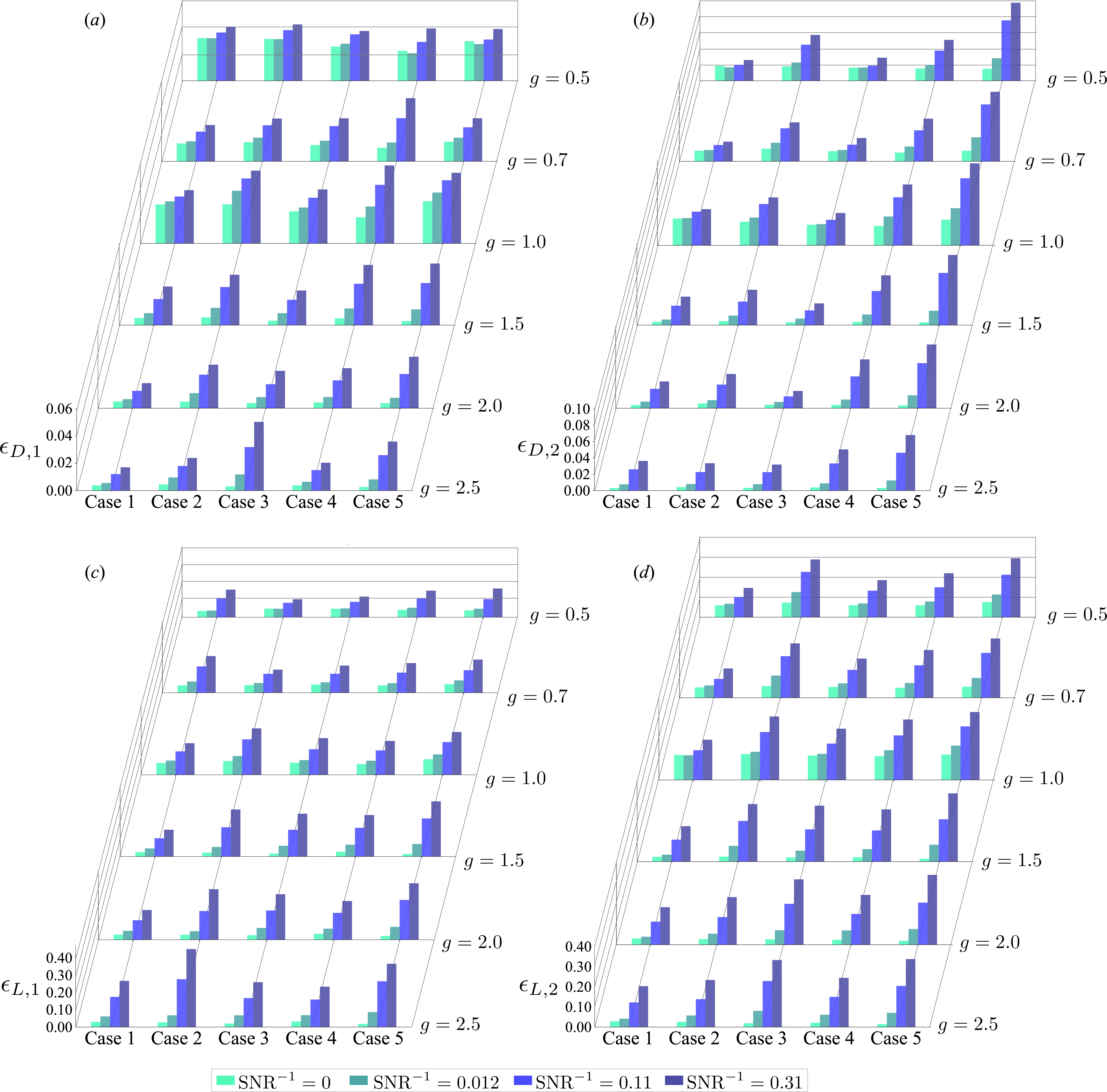}
	\caption{{\color{red}Robustness against noisy input for a flow over two side-by-side cylinders with its radius ratio of $r=1.30$.}}
	\label{fig:A3}
\end{figure*}

{\color{red}We provide the detailed information for the problem presented in figure \ref{fig:twocy_noise}, i.e., the robustness of the CNN-MLP model for noisy inputs.
Figure~\ref{fig:twocy_noise} shows the ensemble $L_2$ error norm averaged over the coefficients of both cylinders for all considered distance factor $g$, in order to show the general trend of the model's response.
In figures~\ref{fig:A1}--\ref{fig:A3}, the result of each flow configurations and coefficients set are presented individually.
From these results too, we can confirm that inserting scalar values at the earlier layer generally leads to the better solution.
}

\bibliographystyle{unsrt}
\bibliography{MFZNF2021}  


\end{document}